\documentclass[lettersize,journal]{IEEEtran}
\usepackage{amsmath,amssymb,amsfonts}
\usepackage{algorithmic}
\usepackage{algorithm}
\usepackage{array}
\usepackage[caption=false,font=normalsize,labelfont=sf,textfont=sf]{subfig}
\usepackage{textcomp}
\usepackage{stfloats}
\usepackage{url}
\usepackage{verbatim}
\usepackage{graphicx}
\graphicspath{{./images/}}
\usepackage{cite}

\usepackage{multirow} % 引入 multirow 包
\usepackage{xcolor}

\newtheorem{theorem}{Theorem}

\def\BibTeX{{\rm B\kern-.05em{\sc i\kern-.025em b}\kern-.08em
		T\kern-.1667em\lower.7ex\hbox{E}\kern-.125emX}}

\begin{document}

%\title{Efficient Container-based Resource Management for Edge Computing with Heterogeneous Tasks}
\title{Squeezing Edge Performance: A Sensitivity-Aware Container Management for Heterogeneous Tasks}

\author{Yongmin Zhang,~\IEEEmembership{Senior Member,~IEEE}, Pengyu Huang,  Mingyi Dong, Jing Yao  \\
	
	\IEEEcompsocitemizethanks{
		\IEEEcompsocthanksitem Y. Zhang,  P. Huang, M. Dong, and J. Yao are with the School of Computer Science and Engineering, Central South University, Changsha, 410083, China. E-mails:\{zhangyongmin, huangpengyu, dongmingyi, jingyao\}@csu.edu.cn. }}

\maketitle

\begin{abstract}
%As an efficient framework to provide low-latency,  high-quality computing services to mobile devices, edge computing has been widely studied in the past decades.  However,  with the explosive growth of heterogeneous tasks, it is difficult for edge computing to satisfy various computational demands due to its limited resources and inflexible resource management architecture. To address this dilemma, we propose a container-based resource management framework for edge servers, where heterogeneous tasks can be processed efficiently by adjusting the container configuration adaptively. We firstly investigate the relationship between type heterogeneity and resource demands of tasks and model a correlation function of task processing delay and container configuration, including CPU and memory, from mass testing data to characterize the resource intensity of heterogeneous tasks.  Then, considering both the energy consumption and the processing delay, we formulate a container-based resource management optimization problem, which is a mixed integer nonlinear programming problem (MINLP), to explore the optimal number of containers and their corresponding resource configurations.  To solve this problem, we prove that the MINLP can be transformed into a solvable convex optimization problem, and propose an efficient container-based resource management scheme (CRMS) to optimize the energy consumption and the processing delay simultaneously.  Finally,  extensive simulation results demonstrate the efficiency of the proposed CRMS compared to existing ones.

Edge computing enables latency-critical applications to process data close to end devices, yet task heterogeneity and limited resources pose significant challenges to efficient orchestration. This paper presents a measurement-driven, container-based resource management framework for intra-node optimization on a single edge server hosting multiple heterogeneous applications. Extensive profiling experiments are conducted to derive a nonlinear fitting model that characterizes the relationship among CPU/memory allocations and processing latency across diverse workloads, enabling reliable estimation of performance under varying configurations and providing quantitative support for subsequent optimization. Using this model and a queueing-based delay formulation, we formulate a mixed-integer nonlinear programming (MINLP) problem to jointly minimize system latency and power consumption, which is shown to be NP-hard. The problem is decomposed into tractable convex subproblems and solved through a two-stage container-based resource management scheme (CRMS) combining convex optimization and greedy refinement. The proposed scheme achieves polynomial-time complexity and supports quasi-dynamic execution under global resource constraints. Simulation results demonstrate that CRMS reduces latency by over 14\% and improves energy efficiency compared with heuristic and search-based baselines, offering a practical and scalable solution for heterogeneous edge environments with dynamic workload characteristics.
\end{abstract}


\section{Introduction}
\label{sec:intro}
The rapid proliferation of Internet of Things (IoT) technologies has driven the deployment of latency-sensitive applications, including augmented reality (AR), autonomous driving, and smart healthcare \cite{9462377}. These applications require stringent Quality of Service (QoS) guarantees, such as ultra-low latency, high reliability, and energy efficiency. However, conventional cloud computing architectures based on centralized processing often fail to satisfy these requirements due to significant backhaul transmission delay and limited scalability \cite{8894371}. To address these limitations, edge computing has emerged as a promising paradigm, placing computational resources in close proximity to data sources. By reducing data transmission delays and enabling localized computation, edge computing significantly enhances QoS metrics, positioning it as a critical enabler for modern intelligent systems\cite{long2024efficient}.

Despite its advantages, edge computing faces notable challenges in resource optimization due to the limited computational and storage capacities of individual edge servers. Existing research has primarily addressed network-wide resource utilization through inter-server collaboration strategies, such as task offloading, load balancing, and predictive scheduling \cite{rossi2022dynamic,9763051,9314906,cai2025joint,li2024multi}. While these methods enhance coordination among distributed nodes, they often overlook the problem of how to orchestrate resources within a single resource-constrained edge server hosting multiple applications. In such settings, naive or static policies can assign disproportionate CPU and memory resources to latency-insensitive workloads, leaving latency-critical tasks underprovisioned and degrading overall QoS \cite{9316678,shen2025reinforcement}. This mismatch motivates a resource management framework that not only preserves service quality but also squeezes as much performance as possible out of limited edge resources by reallocating CPU and memory according to task-specific resource sensitivities.

A central challenge in edge resource optimization lies in task heterogeneity, which refers to the diverse resource requirements and sensitivities of tasks to computational resources. Unlike homogeneous workloads, edge applications can differ drastically in how their latency responds to changes in CPU, memory, and concurrency levels, even when running on the same hardware and container runtime environment \cite{9316678}. For example, memory-intensive workloads may experience significant performance degradation when memory is reduced, whereas some compute-oriented models are more resilient under similar conditions \cite{9745059}. In this paper, we mainly focus on heterogeneity in the CPU and memory demands and sensitivities of applications under a given hardware platform. We refer to these differences in how latency responds to resource adjustments as the resource sensitivity of an application. Without sensitivity-aware allocation, misconfigurations can easily lead to performance bottlenecks, increased energy consumption, and degraded QoS.

Moreover, task heterogeneity also interacts with system-level constraints in a nontrivial manner. When multiple containerized applications share a single edge server, the limited CPU and memory resources couple their performance: improving the latency of one application often requires reducing resources for others. As a result, system performance depends critically on how container quotas are chosen under shared resource constraints, and naive utilization-based scaling or single-resource heuristics often fail to achieve balanced efficiency across heterogeneous workloads. Our measurement-driven fitting results in Section~\ref{sec:fitting} further show that the latency response to CPU and memory adjustments is highly nonlinear and application dependent, indicating that simple threshold rules cannot adequately capture the underlying behavior. These observations motivate a sensitivity-aware container management strategy that exploits heterogeneity in delay–resource elasticity, reallocating CPU and memory from low-elasticity workloads to high-elasticity ones to squeeze more performance from a resource-constrained edge server.

%To address these issues, this paper presents an in-depth study on dynamic container-based resource management for resource-constrained edge servers. This work focuses on intra-node resource orchestration for a single edge server that hosts multiple container clusters, where the goal is to jointly configure the number of containers and their CPU and memory quotas for co-located applications under global resource constraints. We conduct extensive profiling experiments on representative compute-intensive stream applications using Docker containers configured with varying levels of CPU and memory. From the experimental data, we derive a measurement-driven fitting function that models the relationship among task type, processing latency, and container resource configuration. Based on this function and a queueing-based delay model, we formulate a mixed-integer nonlinear programming problem that jointly optimizes latency and energy consumption. To solve this problem efficiently, we design a sensitivity-aware container-based resource management scheme (CRMS) that reallocates CPU and memory from less sensitive applications to more sensitive ones while respecting shared resource resource constraints. Our main contributions can be summarized as follows:
To address these issues, this paper presents a dynamic container-based resource management framework for a single resource-constrained edge server that hosts multiple container clusters. The goal is to jointly determine the number of containers and their CPU and memory quotas under shared resource constraints. Based on extensive profiling across heterogeneous applications, we derive a measurement-driven performance model capturing the nonlinear relationship among resource allocations and processing latency. Leveraging this model, we formulate a mixed-integer nonlinear optimization problem to minimize both latency and energy consumption, and design a sensitivity-aware resource management scheme to achieve efficient intra-node resource orchestration. Our main contributions are summarized as follows:
\begin{itemize}
	\item We construct a container-based fitting function that captures the relationship between processing delay and resource configurations (CPU and memory) for heterogeneous tasks, and use it to characterize the heterogeneous resource sensitivities of different applications.
	\item We formulate the resource management problem as a joint optimization of processing latency and energy consumption, modeled as a MINLP problem, and propose a sensitivity-aware container-based resource management scheme (CRMS) to derive near-optimal allocation strategies under latency–energy trade-offs.
	\item Extensive simulations show that the proposed scheme effectively adapts to diverse workload patterns, reducing average response latency by at least 14\% compared to traditional search algorithms.
\end{itemize}

%The remainder of this paper is organized as follows. Section~\ref{sec:related} reviews related work on edge resource management and performance modeling for heterogeneous applications. Section~\ref{sec:fitting} introduces the container-based testing methodology and the fitting function. Section~\ref{sec:system_model} presents the system model, including the power model, the queueing-based delay model, and the formal problem formulation. Section~\ref{sec:theorem_algorithm} describes the proposed server resource management scheme and the derivation of optimal container configurations based on theoretical analysis. Section~\ref{sec:simulation} reports simulation results to evaluate the effectiveness and robustness of the scheme. Finally, Section~\ref{sec:conclusion} concludes the paper.

The remainder of this paper is organized as follows. Section~\ref{sec:related} reviews related works. Section~\ref{sec:fitting} presents the measurement-driven container-based performance model and profiling methodology. Section~\ref{sec:system_model} introduces the system architecture, power consumption model, processing delay model, and formulates the resource management optimization problem. Section~\ref{sec:theorem_algorithm} describes the proposed server resource management scheme and the derivation of optimal container configurations based on theoretical analysis. Section~\ref{sec:simulation} provides extensive simulation results to evaluate the scheme’s effectiveness. Finally, Section~\ref{sec:conclusion} concludes the paper.

\section{Related Work}
\label{sec:related}
Efficient resource allocation is fundamental to ensuring service quality on resource-constrained edge servers. A wide spectrum of approaches has been investigated—from heuristic autoscaling and analytical modeling to learning-based and serverless frameworks. This section reviews these studies from a methodological perspective, focusing on rule-based analytical models, heterogeneity-aware optimization, and hybrid or serverless architectures. It further highlights their limitations in managing heterogeneous task workloads.

Early studies primarily focused on optimizing resource usage for homogeneous or predictable tasks. Threshold-based methods \cite{rossi2022dynamic,9763051,9314906} adjust CPU or memory allocations when metrics exceed predefined limits. However, such thresholds lack the flexibility needed for tasks that demand simultaneous scaling of multiple resources, often resulting in resource bottlenecks. Queue-theoretic approaches \cite{li2022cost,9764612} model task arrivals and service rates to minimize average response time. While effective for homogeneous tasks, these methods are insufficient for prioritizing latency-critical tasks in heterogeneous environments. Predictive methods \cite{9076292,9328525,9149819} forecast future demands using historical data, but their prediction accuracy significantly degrades when task characteristics and resource demands become unpredictable. These limitations highlight the need for adaptability and sensitive-aware strategies in heterogeneous edge environments.

A second line of work explicitly addresses task heterogeneity through dynamic prioritization and resource partitioning. Sharif \emph{et al.} \cite{9534772} proposed A-PBRA, a method that classifies tasks based on resource sensitivity (e.g., CPU-bound vs. memory-bound) and allocates resources accordingly. He \emph{et al.} \cite{10319405} introduced a three-stage model (TSHC) to jointly optimize task offloading and resource allocation under deadline constraints, yet its reliance on a centralized scheduler limits scalability in distributed edge environments. Control-theoretic methods \cite{zhong2022pacc,9442296,9512507} dynamically adjust resources via feedback loops, but their reliance on pre-defined performance models limits flexibility for unseen task types. Reinforcement learning (RL) \cite{pang2022towards,kardani2020adrl} learns allocation policies without relying on explicit modeling, but the high training overhead and slow convergence of RL limit its applicability in real-time scenarios with bursty, heterogeneous workloads.

To balance flexibility with efficiency, hybrid architectures have emerged. Han \emph{et al.} \cite{9641503} decoupled task scheduling from resource provisioning via a hierarchical framework, enabling differentiated handling of tasks. Ascigil \emph{et al.} \cite{9326369} compared centralized versus decentralized allocation, showing that partial coordination improves deadline compliance for mixed-criticality tasks. In serverless edge environments, Raza \emph{et al.} \cite{raza2023configuration} used Bayesian optimization to tune resource configurations per function type, while Safaryan \emph{et al.} \cite{safaryan2022slam} optimized memory allocations by tracing inter-function dependencies. However, these solutions optimize isolated resources or assume predictable task workflows, neglecting the necessity of coordinated multi-resource adaptation.

Overall, existing approaches either optimize isolated metrics, rely on static performance models, or incur substantial online learning costs, which makes them difficult to apply directly to fine-grained CPU and memory quota management for heterogeneous containerized workloads on a single edge node. In this work, we take a complementary approach by focusing on intra-node resource management and developing a lightweight, measurement-driven scheme that builds a container-level performance model and periodically re-optimizes CPU and memory quotas, as well as the number of containers, across heterogeneous workloads, without requiring online learning or frequent retraining.

\section{Container-based Performance Model}
\label{sec:fitting}
In this section, we build a container-based performance model that maps allocated CPU and memory to the average processing delay of each application. This empirical latency--resource function provides a foundation for the subsequent processing delay model and resource optimization.

\subsection{Modeling Motivation}
Docker containers have been widely deployed on edge servers to process heterogeneous tasks, thanks to their advantages, including short startup time, low memory overhead for encapsulating applications, and elastic scalability\cite{dogani2024proactive}. As a result, dynamic resource management in edge computing can be achieved by controlling the resource configurations of Docker containers on the edge server. Most existing research related to Docker containers\cite{shi2023auto,zhang2021joint,dogani2024proactive} assumes that resource allocations are fixed or that processing rates scale linearly with task size. However, without considering the heterogeneity and resource demands of tasks, this assumption may not hold for edge servers processing heterogeneous workloads.

To verify this assumption, we consider image classification and object detection as two representative categories of heterogeneous tasks. By deploying various models from these two categories in Docker containers, we conduct extensive profiling experiments to examine how processing latency varies with different CPU and memory configurations. The resulting measurements serve as the foundation for constructing the container-based performance model presented in this section.

\subsection{Experimental Setup}
%In this paper, we use container image files provided by the Baidu PaddlePaddle platform to implement container deployment and conduct corresponding experiments. The image classification application is built by ResNet\_v2, SE\_ResNeXt, and MobileNet\_v2 models, respectively, and the object detection application is built with SSD\_MobileNet\_v1. Besides, to achieve resource management through Docker Swarm tools, we set the "--cpus" startup parameter to the CPU share that can be used for the container. As for memory resources, the "-m" startup parameter is set to the maximum memory available for the container. By sending image processing requests to the application containers continuously, we adopt the iteration method to record the processing delay of different application containers.  The iteration method means that,  if one type of resource (CPU or memory) is not limited, the other can be adjusted in a certain granularity for experiments. 
The experiments are conducted using Docker Swarm as the container orchestration tool. Image processing applications are deployed with PaddlePaddle official images to simulate heterogeneous tasks. Four representative applications are tested: ResNet\_v2, SE\_ResNeXt, and MobileNet\_v2 for image classification, and SSD\_MobileNet\_v1 for object detection. 
The CPU and memory resources are controlled by the Docker parameters \texttt{--cpus} and \texttt{-m}, which specify the maximum CPU quota and memory allocation, respectively. 

By continuously sending image processing requests to the application containers, we measure the average processing delay under each resource configuration. In each experiment series, we fix one resource (CPU or memory) to a sufficiently large value and vary the other over a set of discrete configurations, so that the individual impact of CPU and memory on processing latency can be characterized separately.

\begin{figure}[h]
	\centering
	\begin{minipage}[t]{0.49\textwidth}
		\centering
		\includegraphics[width=\textwidth]{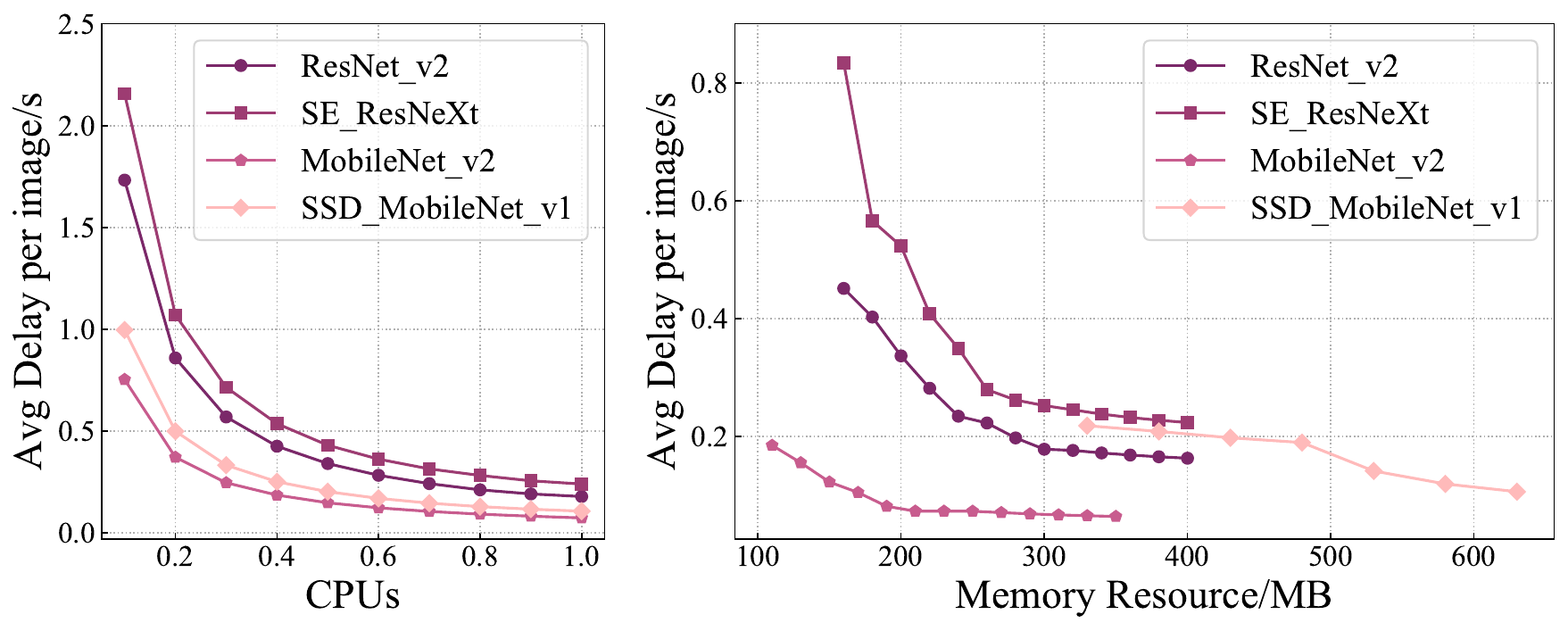}
	\end{minipage}
	\vspace{-18pt}
	\caption{Impact of CPU and memory allocations on processing delay.}
	\label{Test}
	\vspace{-1em}
\end{figure}
\subsection{Profiling Observations}
%add
We control container resources via Docker runtime parameters. Specifically, \texttt{--cpus} sets a CPU time \emph{limit} via the CFS quota/period mechanism (allowing fractional capacity, e.g., \texttt{--cpus=1.5}), while \texttt{-m} sets a hard memory limit. In our profiling, we vary these two parameters to obtain latency measurements under different CPU and memory configurations.

\begin{table*}[htb]
	\centering
	\caption{RMSE and R-Square comparison of candidate fitting functions.}
	\begin{tabular}{|c|c|c|c|c|c|c|c|c|}
		\hline
		\multirow{2}{*}{Function Model} & \multicolumn{2}{c|}{ResNet\_v2} & \multicolumn{2}{c|}{SE\_ResNeXt} & \multicolumn{2}{c|}{MobileNet\_v2} & \multicolumn{2}{c|}{SSD\_MobileNet\_v1} \\
		\cline{2-9}
		& RMSE & R-Square & RMSE & R-Square & RMSE & R-Square & RMSE & R-Square \\
		\hline
		$\frac{\kappa_1}{1-e^{-\kappa_2 r^{\text{cpu}}}}+e^{\frac{\kappa_3}{r^{\text{mem}}}}$ & 2.69 &  0.99 & 4.17 & 0.99 & 0.72 & 0.99 & 2.79 & 0.99 \\
		\hline
		$\frac{\kappa_1}{r^{\text{cpu}}}+\kappa_2 \left(r^{\text{mem}}\right)^2+\kappa_3 r^{\text{mem}}$ & 4.17 & 0.99 & 5.77 & 0.99 & 1.94 & 0.99 & 1.95 & 0.97 \\
		\hline
		$\frac{1}{\kappa_1 \log \left(1+r^{\text{cpu}}\right) + \kappa_2 \log \left(1+r^{\text{mem}}\right)}$ & 7.41 & 0.97 & 12.10 & 0.95 & 3.32 & 0.97 & 4.84 & 0.97 \\
		\hline
		$\frac{\kappa_1}{\kappa_2 + \kappa_3 \left(r^{\text{cpu}}\right)^2 + \kappa_4 \left(r^{\text{mem}}\right)^2}$ & 14.88 & 0.89 & 21.84 & 0.85 & 6.15 & 0.91 & 8.61 & 0.91 \\
		\hline
		$\kappa_1 \left(r^{\text{cpu}}\right)^3 + \kappa_2 \left(r^{\text{mem}}\right)^3 + \kappa_3 r^{\text{cpu}} r^{\text{mem}}$ & 23.31 & 0.74 & 28.12 & 0.76 & 10.88 & 0.71 & 15.33 & 0.70 \\
		\hline
	\end{tabular}
	\label{table:fitting}
	\vspace{-1em}
\end{table*}

\begin{figure*}[t]
	\centering
	\begin{minipage}[b]{0.49\textwidth}
		\centering
		\includegraphics[width=0.49\textwidth]{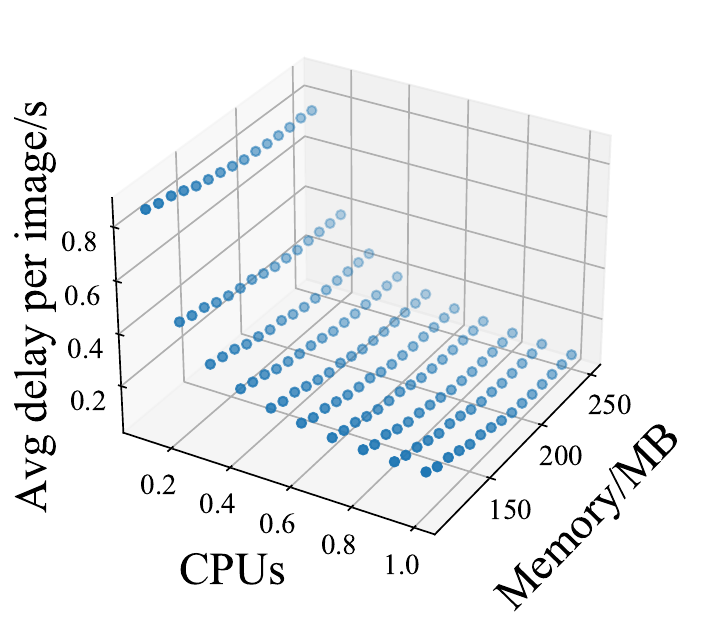}
		\includegraphics[width=0.49\textwidth]{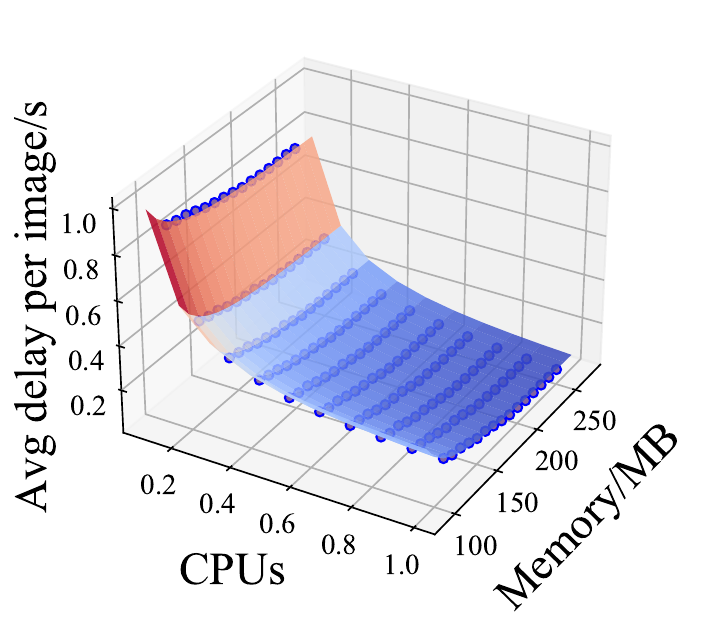}
		\caption{Fitted latency--resource surfaces versus measured data.}
		\label{fitting}
	\end{minipage}
	\hfill
	\begin{minipage}[b]{0.49\textwidth}
		\centering
		\includegraphics[width=0.49\textwidth]{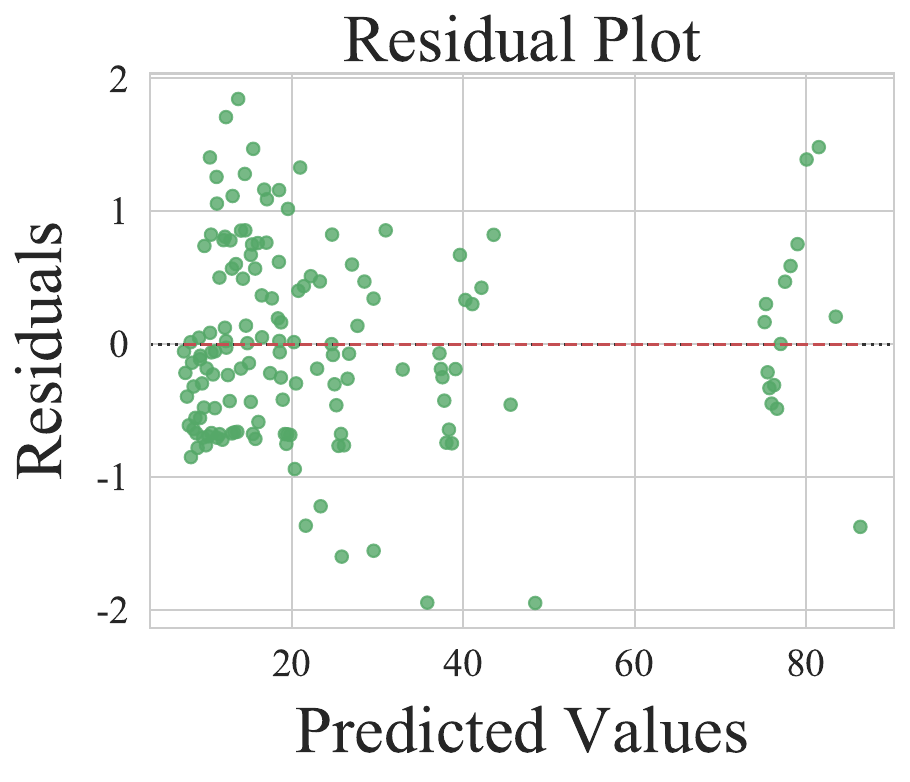}
		\includegraphics[width=0.49\textwidth]{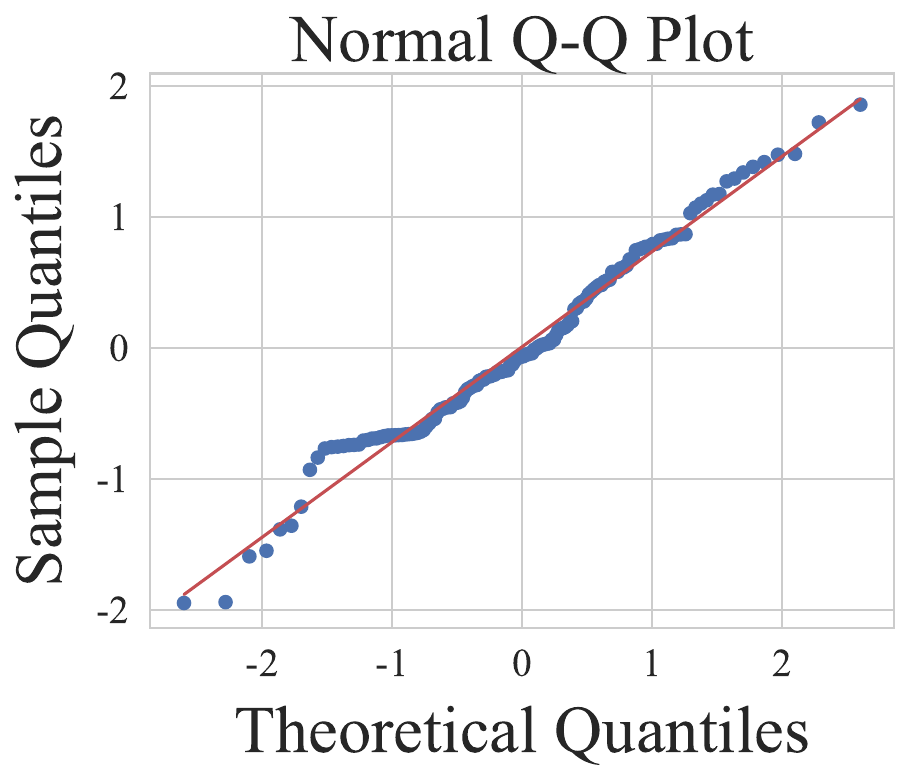}
		\caption{Residual and Q--Q plots verifying the performance model fitting.}
		\label{residual_QQ_Plot}
	\end{minipage}
	\vspace{-1em}
\end{figure*}

When memory resources are unrestricted, we evaluate how processing latency varies with CPU allocation across four applications, as shown in Fig.~\ref{Test}(a). The results reveal a non-linear increase in latency as the available CPU capacity decreases. Moreover, the degree of sensitivity to CPU resources differs markedly among applications: SE\_ResNeXt is most sensitive to CPU reduction, followed by ResNet\_v2, MobileNet\_v2, and SSD\_MobileNet\_v1. These results highlight the importance of fine-grained CPU allocation tailored to each application's computational demand.

Conversely, when CPU resources are unrestricted, we investigate the effect of memory allocation on processing latency for the same four applications, as shown in Fig.~\ref{Test}(b). The SSD\_MobileNet\_v1 object detection model exhibits the highest memory requirement, while MobileNet\_v2 shows a similar but less pronounced trend. In contrast, ResNet\_v2 and SE\_ResNeXt are more sensitive to memory reductions, where even small decreases lead to sharp increases in latency. A clear lower bound is also observed: when memory allocation falls below a threshold, the container terminates due to an Out-of-Memory (OOM) error. This empirical observation underscores that both CPU and memory are performance-critical resources, and adequate memory provisioning is essential for stable container execution.

Based on these observations, for each application $i$, we define a minimum feasible memory $r_i^{\min}$ as the smallest allocation that avoids OOM events, and a saturation point $r_i^{\max}$ beyond which additional memory yields negligible latency improvement. These empirically derived bounds are later used to constrain the feasible region in the optimization model. Fig.~\ref{Test} illustrates that heterogeneous tasks have distinct demand characteristics for CPU and memory resources, which contradicts the assumption that the processing rate is a linear function of task size. To further explore the relationship between processing delay and resource configuration, i.e., CPU and memory, we further construct a container-level performance model of processing latency as a function of CPU and memory resources in the next subsection. 

\subsection{Model Fitting and Validation}
We conduct latency profiling for four applications under various CPU and memory configurations. To reduce noise and measurement errors, each data point represents the average of multiple valid runs, yielding stable results for curve fitting. Based on the collected data, we fit a container-level latency--resource function $d_i(r_i^{cpu}, r_i^{mem})$ for each application $i$. Using the SciPy toolkit, we perform non-linear least squares fitting across five candidate models, evaluated by the Root Mean Square Error (RMSE). A smaller RMSE indicates a closer match between predicted and measured latency. The RMSE and $R^2$ values for all candidates are listed in Table~\ref{table:fitting}, from which Eq.~(\ref{eq.containerD}) achieves the best fit and is therefore adopted to model the average processing delay of application $i$:

\begin{equation} \label{eq.containerD}
	d_i=\frac{\kappa_{i,1}}{1-e^{-\kappa_{i,2} r_i^{cpu}}}+ e^{\frac{\kappa_{i,3}}{r_i^{mem}}},
\end{equation}
where $r_i^{cpu}$ and $r_i^{mem}$ denote the CPU and memory resources allocated to the container for application $i$. The parameters $\kappa_{i,1}<0$, $\kappa_{i,2}>0$, and $\kappa_{i,3}>0$ are the fitted coefficients obtained from the non-linear regression. These coefficients capture each application's sensitivity to CPU and memory allocations. Intuitively, a larger $\kappa_{i,2}$ makes the first term more responsive to changes in $r_i^{cpu}$, while a larger $\kappa_{i,3}$ amplifies the effect of $r_i^{mem}$ in the exponential term. The subsequent resource management scheme exploits these heterogeneous sensitivities when redistributing CPU and memory under global constraints.

To validate the suitability of Eq.(\ref{eq.containerD}), we compare its fitted latency surface with actual measurements from the MobileNet\_v2-based image classification application under varying CPU and memory configurations. The scatter points align closely with the model's predicted surface, indicating that the fitting function in Eq.(\ref{eq.containerD}) accurately captures the observed latency trends. This close alignment supports the choice of Eq.(\ref{eq.containerD}) as the performance model to characterize the relationship between processing delay, CPU, and memory resources. This initial observation based on RMSE suggests that the model is relatively reasonable.

To further assess the fitting quality of Eq.~(\ref{eq.containerD}) for the MobileNet\_v2 model, we examine the residual and Q–Q plots shown in Fig.~\ref{residual_QQ_Plot}, which offer additional insights into the model’s accuracy. The residual plot demonstrates that the residuals are randomly distributed, indicating a good fit without systematic bias. The QQ plot reveals that the residuals align closely with the 45-degree reference line, suggesting that they approximate a normal distribution, which is desirable for model accuracy. Additionally, the adjusted R-squared value is exceptionally high at 0.9987, indicating that the model accounts for nearly all the variability in the data. The mean squared error (MSE) of 0.5568 and root mean squared error (RMSE) of 0.7223 further reflect the model's high predictive accuracy, underscoring its robustness in capturing the relationship between CPU/memory configurations and inference latency. Collectively, these metrics confirm that Eq. (\ref{eq.containerD}) is an effective and reliable choice for modeling the inference latency, providing a strong foundation for further optimization and analysis.

\section{System Model}
\label{sec:system_model}
\begin{figure}[hbtp]
	\centering
	\begin{minipage}[b]{0.4\textwidth}
		\centering
		\includegraphics[width=\textwidth]{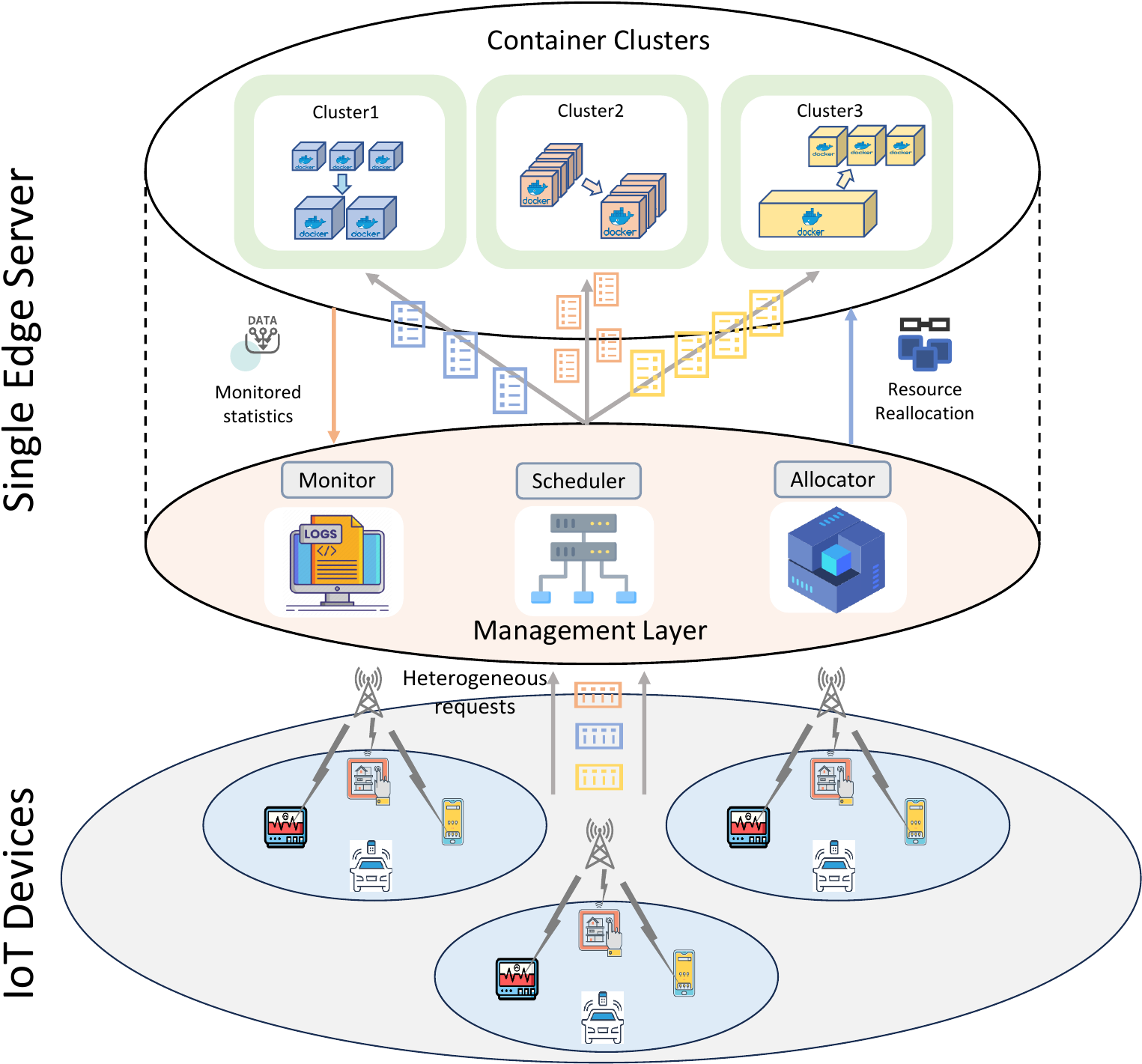}
	\end{minipage}
	%	\vspace{-6pt}
	\caption{Architecture of the container-based resource management framework within a single edge server.}
	\label{framework1}
	\vspace{-1em}
\end{figure}

In this section, we present a container-based resource management framework for a single edge server, as illustrated in Fig.~\ref{framework1}. The server hosts multiple containerized applications with heterogeneous workload characteristics. These include distinct request arrival rates, varying average request sizes (measured by the number of images per request), and diverse latency–resource sensitivities as modeled in Section~\ref{sec:fitting}. The framework focuses on efficient intra-node management of CPU and memory resources within Docker container clusters by coordinating monitoring, scheduling, and allocation.

IoT devices, including sensors, traffic cameras, and smartphones, generate requests that are transmitted over wireless links to the edge server for computation. Let $\mathcal{M}=\{1,2,\dots,m\}$ denote the set of applications hosted on the server. For application $i\in\mathcal{M}$, the edge server maintains a container cluster with $N_i$ homogeneous containers. Each incoming request of application $i$ is processed by one container instance. Within each container, requests follow a First-Come-First-Serve (FCFS) scheduling policy. The average request size of application $i$, defined as the mean number of images per request, will later be used to relate the service rate to the per-image latency.

The system is logically divided into a management plane and a worker plane. The worker plane runs the container clusters and executes the application requests. The management plane consists of three functional modules: the monitor, scheduler, and the allocator, which cooperate to achieve dynamic resource management. The monitor continuously collects runtime statistics, including request arrival rates, average processing delays, and container resource configurations, to capture significant workload variations. Using the monitoring information, the scheduler assigns incoming requests to the corresponding application clusters according to their types, ensuring proper task routing without altering resource quotas. The allocator updates container configurations in response to workload variations detected by the monitor: it performs horizontal scaling by adjusting the number of containers $N_i$ and vertical scaling by adjusting the CPU and memory quotas $(r_i^{cpu}, r_i^{mem})$ of each application. The exact triggering policy of resource reconfiguration (periodic or on-demand) is implementation-dependent and not the focus of this study; here we assume the allocator reacts when the monitor reports material workload changes.

To evaluate and optimize system performance, two quantitative models are introduced in the following subsections: a power consumption model and a processing delay model. The former relates container resource allocations to the incremental power attributable to applications at the edge server, while the latter estimates the per-image latency of each application given its container-level CPU and memory configuration. These models provide the foundation for the resource management optimization problem developed subsequently.

\subsection{Power Consumption Model}
Different from existing power models that parameterize server power as a function of CPU frequency and chip-level energy coefficients, we model power as a function of the allocated CPU capacity. This design choice aligns with container orchestration mechanisms, where the adjustable resource control is the CPU quota rather than processor frequency. Following the modeling principle in \cite{9448450}, the dynamic power consumption is approximated as a linear function of the allocated CPU quota. We define the incremental power of application $i$ as:
\begin{equation}
	\Delta P_i = (P^{full}-P^{idle})\, U_i^{cpu},
	\label{eq:power_increment}
\end{equation}
where $P^{idle}$ and $P^{full}$ are the measured server power at idle and at a reference full-load state, respectively. The server power is thus expressed as $P^{idle} + \sum_i \Delta P_i$, and the constant idle term $P^{idle}$ does not affect the optimization objective, so it is omitted in subsequent analysis. Here $U_i^{cpu}$ denotes the fraction of total CPU capacity allocated to application $i$:
\begin{equation}
	U_i^{cpu} = \frac{N_i\, r_i^{cpu}}{\overline{R^{cpu}}},
	\label{eq:Cpu_usage}
\end{equation}
where $N_i$ is the number of containers for application $i$, $r_i^{cpu}$ is the CPU quota of each container, and $\overline{R^{cpu}}$ is the total CPU capacity of the edge server. This formulation uses the capacity fraction $U_i^{cpu}$ as a practical proxy for the workload-proportional power term, aligning the model with the resource control interface exposed by the container runtime while remaining consistent in spirit with utilization-based power modeling. This model focuses on CPU power, while memory power is not explicitly modeled. In typical edge servers, memory power exhibits much smaller variation compared to that of the CPU under our workload settings. Its effect can therefore be absorbed into the constant terms $P^{idle}$ and $P^{full}$ without loss of generality. Extending the model to include explicit memory power will be considered in future work.

\subsection{Processing Delay Model}
We model the request processing of each application as an M/M/N queue, where request arrivals follow a Poisson process with rate $\lambda_i$ and service times in each container are exponentially distributed with rate $\mu_i$. The expected number of requests in the system (including both the waiting queue and the servers), denoted by $L_s(N_i,\lambda_i,\mu_i)$, is given by
\begin{eqnarray}\label{eq:length}
	&&L_s(N_i, \lambda_i, \mu_i)=\frac{\left(\frac{\lambda_i}{\mu_i}\right)^{N_i} \frac{\lambda_i}{N_i \times \mu_i}}{N_i !\left(1-\frac{\lambda_i}{N_i \times \mu_i}\right)^2} \,\pi_0+\frac{\lambda_i}{\mu_i},\\
	&&\pi_0=\left[\sum_{k=0}^{N_i-1} \frac{1}{k !}\left(\frac{\lambda_i}{\mu_i}\right)^k+\frac{\lambda_i^{N_i}}{N_i !\left(1-\frac{\lambda_i}{N_i \times \mu_i}\right) \mu_i^{N_i}}\right]^{-1} .
\end{eqnarray}
Here, $\pi_0$ denotes the probability that there are no requests in the system, and $\mu_i$ is the per-container service rate for application $i$. Considering the latency–resource relationship in Eq.~(\ref{eq.containerD}), the service rate can be written as:
\begin{equation}\label{eq.uxd}
	\mu_i =\frac{1}{\bar{x_i} d_i},
\end{equation} 
where $d_i$ is the average \emph{per-image} processing latency under the allocated CPU and memory, and $\bar{x_i}>0$ denotes the average number of images contained in one request of application $i$; hence $\bar{x_i} d_i$ is the average processing time per request. According to Little’s law, the expected average response time per request for application $i$, denoted by $W_s(N_i,\lambda_i,\mu_i)$, can be given by
\begin{equation} \label{Eq.mmn}
	W_s(N_i,\lambda_i,\mu_i)=\frac{L_s(N_i,\lambda_i,\mu_i)}{\lambda_i}.
\end{equation}

\subsection{Problem formulation}
Based on the preceding models, we formulate a container-based resource management optimization problem, which determines the optimal number of containers and their CPU and memory configurations for each application. The objective is to jointly minimize the processing delay and power consumption under resource capacity constraints. The problem is formulated as follows: 
\begin{flalign}
	\textbf{\text {P: }} \min_{N_i,r_{i}^{cpu},r_{i}^{mem}}~ & U_p=\sum\limits_{i}\!\left(\alpha\, W_s(N_i,\lambda_i,\mu_i)+\beta\, \frac{\Delta P_i}{\lambda_i}\right)\label{eq:p}, \\
	\text { s.t. }\qquad & \sum\limits_{i}{N_{i} r_{i}^{cpu}} \le \overline{R^{cpu}}, \label{ys.p.1}\\
	~& \sum\limits_{i}{N_{i} r_{i}^{mem}} \le \overline{R^{mem}}, \label{ys.p.2}\\
	~& r_{i}^{min}\le r_{i}^{mem}\le r_{i}^{max},~ \forall i \in M. \label{ys.p.3}
\end{flalign}

Here, $\alpha>0$ and $\beta>0$ are the weight coefficients for delay and energy terms, respectively. $W_s(N_i,\lambda_i,\mu_i)$ denotes the average response time per request of application $i$, while $\Delta P_i/\lambda_i$ represents the average incremental power consumption per request. Therefore, $U_p$ minimizes the weighted sum of per-request delay and per-request energy across all applications. Constraints~(\ref{ys.p.1}) and~(\ref{ys.p.2}) ensure that the total CPU and memory allocations do not exceed the available system capacities $\overline{R^{cpu}}$ and $\overline{R^{mem}}$, respectively. Constraint~(\ref{ys.p.3}) bounds the memory assigned to each container, preventing insufficient memory that could trigger the Out-of-Memory (OOM) killer and avoiding excessive memory allocations that starve other applications. 

In Problem~\textbf{P}, $N_i$ is a non-negative integer variable, while $r_{i}^{cpu}$ and $r_{i}^{mem}$ are continuous decision variables, making it a mixed-integer nonlinear programming (MINLP) problem~\cite{9555607}. To further characterize its computational complexity, we next present a polynomial-time reduction from the well-known Unbounded Multi-dimensional Knapsack Problem (UMKP), which is NP-hard~\cite{kellerer2004multidimensional}, leading to the following theorem.

\begin{theorem}
	Problem \textbf{P} is NP-hard.
\end{theorem}
\begin{IEEEproof}
	We prove this by reduction from the Unbounded Multi-dimensional Knapsack Problem. Let $k_i$ denote the number of selected items of type $i$ in the UMKP, which is formulated as
	\begin{equation}
		\max \sum_{i=1}^{m} k_i v_i,\quad
		\text{s.t.}\quad \sum_{i=1}^{m} k_i w_i^{(j)} \le C^{(j)},\ \forall j\!\in\![1,d].
		\label{eq:umkp}
	\end{equation}
	Here, $v_i$ and $w_i^{(j)}$ denote the profit and the $j$-th resource weight of item $i$, respectively, and $C^{(j)}$ is the capacity of dimension $j$.
	
	We construct a special case of Problem~\textbf{P} as follows. Each application $i$ corresponds to an item type $i$ in the UMKP, and the number of deployed containers $N_i$ corresponds to the number of selected items $k_i$. The CPU and memory allocations map to the two-dimensional resource weights, i.e., $r_i^{cpu}=w_i^{(1)}$ and $r_i^{mem}=w_i^{(2)}$, with total capacities $\overline{R^{cpu}}=C^{(1)}$ and $\overline{R^{mem}}=C^{(2)}$. For the objective function, we simplify Problem~\textbf{P} by setting $\alpha=0$ and $\beta=1$, thereby eliminating the delay term and focusing only on the energy term. Assuming a linear incremental power model $\Delta P_i = c_i N_i$, where $c_i>0$ is a constant, the objective becomes:
	\begin{equation}
		\min \sum_{i} \frac{c_i N_i}{\lambda_i}.
	\end{equation}
	Since $\lambda_i$ is the arrival rate of requests for application $i$, $\frac{c_i N_i}{\lambda_i}$ can be interpreted as the average incremental energy cost per request under this simplified model. To align this minimization problem with the profit-maximization objective of the UMKP, we define $\frac{c_i}{\lambda_i}=-v_i$. Then, we obtain:
	\[
	\min \sum_i \frac{c_i N_i}{\lambda_i}
	= \min \sum_i (-v_i) N_i
	= -\max \sum_i v_i N_i.
	\]
	Thus, an optimal solution to this special case of Problem~\textbf{P} directly yields an optimal solution to the corresponding UMKP instance. Because UMKP is NP-hard~\cite{kellerer2004multidimensional}, Problem~\textbf{P} is at least NP-hard. Moreover, reintroducing the nonlinear queuing delay term $W_s(N_i,\lambda_i,\mu_i)$ with factorial and exponential expressions further increases the complexity, making the problem non-convex and computationally intractable~\cite{floudas2013deterministic}.
\end{IEEEproof}

Given the NP-hardness of Problem P, exact solutions become computationally infeasible for large-scale instances. To address this challenge, we conduct a problem transformation through theoretical analysis and subsequently develop an efficient resource management algorithm tailored to this setting.

\section{Sensitivity-aware Container Resource Management Scheme for Heterogeneous Tasks}
\label{sec:theorem_algorithm}
In this section, we transform Problem~\textbf{P} into tractable subproblems through theoretical analysis and design a sensitivity-aware container management scheme (CRMS) for intra-node resource orchestration on a single edge server.

\subsection{Problem Transformation in Sufficient Resources Condition}
To handle the coupling between the container count $N_i$ and the per-container resource quotas $(r_i^{cpu}, r_i^{mem})$ in Problem~\textbf{P}, we first analyze per-application optimal configurations under a sufficient-resource assumption, where the global CPU and memory budget constraints in (\ref{ys.p.1})–(\ref{ys.p.2}) are temporarily relaxed. Under this assumption, each application can be optimized independently, leading to two decoupled subproblems: \textbf{SP1} (per-container quota selection) and \textbf{SP2} (container count selection). The resulting configurations characterize the upper-bound resource demand of each application and will later serve as initialization and guidance for the constrained scheme.

%For Subproblem \textbf{SP1}, considering independent processes of containers, the resource allocation of each container is independent when the edge server has sufficient resources. Thus, we can derive the optimal resource configuration of each container for application $i,\forall i\in M$ in Subproblem \textbf{SP1}, 
%\begin{flalign}
%	\textbf{\text { SP1: }} \min_{r_i^{cpu}, r_i^{mem}}~ & F_{i}=\alpha D_{i}+\beta \frac{p_{i}}{\lambda_{i}}\label{eq:Fi}, \\
%	\text { s.t. }~ & D_i=d_i  \bar{x_i}, \\
%	~& r_{i}^{min}\le r_{i}^{mem}\le r_{i}^{max},
%\end{flalign}
%where $F_i$ represents the utility for processing delay and power consumption of application $i$, $D_i$ represents the processing delay of the container for each task, and $p_{i}$ denotes the power consumption of a container for application $i$. The value of $p_{i}$ can be given by
%\begin{equation}
%	{{p}_{i}}={{P}^{idle}}+({{P}^{full}}-{{P}^{idle}})\centerdot \frac{r_{i}^{cpu}}{\overline{{{R}^{cpu}}}}.
%\end{equation}

For Subproblem \textbf{SP1}, when the edge server has sufficient resources, the allocation to each container can be treated independently. Thus, for application $i,~\forall i\in M$, we determine the optimal per-container CPU and memory by solving:
\begin{flalign}
	\textbf{\text { SP1: }} \min_{r_i^{cpu}, r_i^{mem}}~ & F_{i}=\alpha D_{i}+\beta \frac{\Delta p_{i}}{\lambda_{i}}\label{eq:Fi}, \\
	\text { s.t. }~ & D_i=d_i \,\bar{x_i}, \\
	~& r_{i}^{min}\le r_{i}^{mem}\le r_{i}^{max},
\end{flalign}
where $D_i$ denotes the average per-request processing delay with $D_i=\bar{x}_i\,d_i$, and $\Delta p_i$ denotes the incremental power consumption of one container for application $i$ relative to the idle baseline:
\begin{equation}
	\Delta p_{i}=(P^{full}-P^{idle})\,\frac{r_{i}^{cpu}}{\overline{R^{cpu}}}.
\end{equation} 

To verify the existence and uniqueness of the optimal solution to Subproblem \textbf{SP1}, we analyze the monotonicity and convexity of $F_i$ with respect to $r_i^{cpu}$ and $r_i^{mem}$, which is given as follows:
\begin{theorem}\label{theorem.(1)}
	$F_i$ is a monotonically decreasing function with respect to $r_i^{mem}$ and a convex function of $r_{i}^{cpu}$ and $r_{i}^{mem}$.
\end{theorem}
\begin{IEEEproof}
	Based on Eq.(\ref{eq:Fi}), the first derivative of $F_i$ with respect to $r_i^{cpu}$ and $r_i^{mem}$ can be given by 
	\begin{equation}
		\frac{\partial F_{i}}{\partial r_{i}^{cpu}}=\frac{\alpha \bar{x}_{i} \kappa_{i,1} \kappa_{i,2} e^{\kappa_{i,2}  r_{i}^{cpu}}}{\left(1-e^{\kappa_{i,2}  r_{i}^{cpu}}\right)^2}+\frac{\beta\left(P^{\text {full}}-P^{idle}\right)}{\lambda_{i}\overline{R^{cpu}}} ,
	\end{equation}
	
	\begin{equation}
		\frac{\partial F_{i}}{\partial r_{i}^{mem}}=-\frac{\alpha \bar{x}_{i} \kappa_{i,3}}{\left(r_{i}^{mem}\right)^2}  e^{\frac{\kappa_{i,3}}{r_{i}^{mem}}}.
	\end{equation}
	
	It can be obtained that $\frac{\partial F_{i}}{\partial r_{i}^{mem}}<0$ holds since $\kappa_{i,3}>0$, which means that $F_i$ is a monotonically decreasing function with respect to $r_i^{mem}$. The second derivative of $F_i$ with respect to $r_i^{cpu}$ and $r_i^{mem}$ can be given by
	\begin{equation}
		\frac{\partial^2 F_{i}}{\partial\left(r_{i}^{cpu}\right)^2}=\frac{\alpha \bar{x}_{i} \kappa_{i,1} \kappa_{i,2}^2 e^{\kappa_{i,2}  r_{i}^{cpu}}\left(1-e^{2 \kappa_{i,2}  r_{i}^{cpu}}\right)}{\left(1-e^{\kappa_{i,2}  r_{i}^{cpu}}\right)^4},
	\end{equation}
	\begin{equation}
		\frac{\partial^2 F_{i}}{\partial\left(r_{i}^{mem}\right)^2}=e^{\frac{\kappa_{i,3}}{r_{i}^{mem}}}  \frac{\alpha \bar{x}_{i} \kappa_{i,3}\left(2 r_{i}^{mem}+\kappa_{i,3}\right)}{\left(r_{i}^{mem}\right)^4},
	\end{equation}
	\begin{equation}
		\frac{\partial^2 F_{i}}{\partial r_{i}^{cpu} \partial r_{i}^{mem}}=\frac{\partial^2 F_{i}}{\partial r_{i}^{mem} \partial r_{i}^{cpu}}=0.
	\end{equation}
	
	Since $r_{i}^{cpu}>0, 1-e^{2 \kappa_{i,2}  r_{i}^{cpu}}<0, \kappa_{i,1}<0, \kappa_{i,2}>0$, we have that $\frac{\partial^2 F_{i}}{\partial\left(r_{i}^{cpu}\right)^2}>0$. Besides, $\frac{\partial^2 F_{i}}{\partial (r_{i}^{mem})^2}>0$ due to $\alpha >0,\kappa_{i,3}>0$ and $r_{i}^{mem}>0$. 
	From Hessian Matrix, we can derive that $\frac{\partial^2 F_{i}}{\partial\left(r_{i}^{cpu}\right)^2}  \frac{\partial^2 F_{i}}{\partial\left(r_{i}^{mem}\right)^2}-\frac{\partial^2 F_{i}}{\partial r_{i}^{cpu} \partial r_{i}^{mem}}  \frac{\partial^2 F_{i}}{\partial r_{i}^{mem} \partial r_{i}^{cpu}}>0$ always hold, which means that $F_{i}$ is a convex function with respect to $r_{i}^{cpu}$ and $r_{i}^{mem}$.
\end{IEEEproof}

According to Theorem~\ref{theorem.(1)}, Subproblem \textbf{SP1} is strictly convex and thus admits a unique optimum; the optimal $(r_{i}^{cpu}, r_{i}^{mem})$ can be obtained by standard solvers (e.g., Scipy.optimize, fmincon). Moreover, the monotonicity of $F_i$ with respect to $r_i^{mem}$ implies that the optimal memory is $r_{i}^{mem}=r_{i}^{max}$ under the sufficient-resource assumption.

As for Subproblem \textbf{SP2}, considering the power consumption of container cluster changes with the increase of container number, the objective is to determine the optimal number of containers for each type of application given the optimal solution of \textbf{SP1}.  Subproblem \textbf{SP2} can be formulated as: 
%\begin{flalign}
%	\textbf{\text { SP2: }} \min_{N_i}~ & \Phi(N_i)=\alpha W_{s}(N_i,\lambda_i,\mu_i^{*})+\beta \frac{P(N_i)}{\lambda_{i}} \label{sp2target}, \\
%	\text { s.t. }~ & W_{s}(N_i,\lambda_i,\mu_i^{*})=\frac{L_{s}(N_i,\lambda_i,\mu_i^{*})}{\lambda_i}, \\
%	~ P(N_i)=&P^{idle}+(P^{full}-P^{idle}) \frac{N_i  {r_i^{cpu}}^{*}}{\overline{R^{cpu}}}\label{c:PNi},
%\end{flalign}
\begin{flalign}
	\textbf{\text { SP2: }} \min_{N_i}~ &
	\Phi(N_i)=\alpha W_{s}(N_i,\lambda_i,\mu_i^{*})
	+\beta \frac{\Delta P(N_i)}{\lambda_{i}}, \label{sp2target} \\[3pt]
	\text { s.t. }~ &
	W_{s}(N_i,\lambda_i,\mu_i^{*})
	=\frac{L_{s}(N_i,\lambda_i,\mu_i^{*})}{\lambda_i}, \\[3pt]
	&
	\Delta P(N_i)
	=(P^{full}-P^{idle}) \frac{N_i  {r_i^{cpu}}^{*}}{\overline{R^{cpu}}} \label{c:PNi},
\end{flalign}
where $\Phi(N_i)$ represents the utility of server for application $i$, $\mu_i^{*}$ denotes the optimal service rate of application container $i$, which can be derived by Eq.(\ref{eq.uxd}) given the optimal solutions of \textbf{SP1}.  $P(N_i)$ represents the power consumption of deploying $N_i$ containers for application $i$.
%$W_{s}(N_i,\lambda_i,\mu_i^{*})$ and $L_{s}(N_i,\lambda_i,\mu_i^{*})$ denote the expected time and task queue length of application $i$, repectively, which can be derived by Eqs.(\ref{eq:length}) and (\ref{Eq.mmn}).

%where $\Phi(N_i)$ is the per-request utility for application $i$, $\mu_i^{*}$ is the service rate induced by the optimal per-container resources from \textbf{SP1} via Eq.(\ref{eq.uxd}), and $\Delta P_i(N_i)$ is the \emph{incremental} power of deploying $N_i$ containers for application $i$ (relative to the idle baseline).%
%\footnote{In (\ref{c:PNi}), we keep the original label for cross-referencing consistency; the power term is expressed in incremental form and the constant $P^{idle}$ does not affect optimality.}
 
To verify the existence of optimal solution for Subproblem \textbf{SP2}, we analyze the convexity of $\Phi(N_i)$ with respect to $N_i$, which is shown as following:

\begin{theorem}\label{theorem.(2)}
	$\Phi(N_i)$ is a convex function with respect to $N_i$.
\end{theorem}

\begin{IEEEproof}
	It is easy to find that the power consumption function $\Delta P_i(N_i)$ is a linear function with respect to $N_i$. Thus, the convexity of $\Phi(N_i)$ depends on that of $W_{s}(N_i,\lambda_i,\mu_i^{*})$.  Based on \cite{dyervalidity}, it has been proved that  $W_{s}(N_i,\lambda_i,\mu_i^{*}) $ is a convex function with respect to $N_i$. Thus, we have that $\Phi(N_i)$ is a convex function with respect $N_i$.
\end{IEEEproof}

%According to Theorem.\ref{theorem.(2)}, Subproblem \textbf{SP2} is a convex optimization problem, which can be solved by the ternary search method. For application container cluster $i$, the upper bound of the container number is the number that the server deploys by using all the CPU and memory resources, that is $\min \{\overline{R^{cpu}}/r^{{cpu}^*},\overline{R^{mem}}/r^{{mem}^*}\}$. Besides, the stability of the queue system should always be satisfied, that is $\lambda_i \le N_i  \mu_i$ holds. Thus, there exists the lower bound of the number of containers denoted by $\left\lceil \frac{\lambda_i}{\mu_i} \right\rceil$, where optimal $\mu_i$ can be derived from the optimal solution of \textbf{SP1}.  Based on this, we design an efficient server resource management to solve Problem \textbf{P} in sufficient resource conditions for $M$ kinds of applications, as shown in Algorithm \ref{singleCluster}.

According to Theorem~\ref{theorem.(2)}, Subproblem \textbf{SP2} is a convex one-dimensional optimization and can be solved efficiently via ternary search. For implementation, we define the search interval as
\[
N_i \in \left[\left\lceil \frac{\lambda_i}{\mu_i^{*}} \right\rceil,\; \min\!\left\{\frac{\overline{R^{cpu}}}{r_{i}^{cpu*}},\, \frac{\overline{R^{mem}}}{r_{i}^{mem*}}\right\}\right],
\]
where the lower bound enforces queue stability ($\lambda_i \le N_i \mu_i^{*}$), and the upper bound is the maximum number of containers the server could host for application $i$ if all CPU and memory resources were devoted to this application. The resulting $N_i^{*}$ thus captures the \emph{ideal} container count under the sufficient-resource assumption and serves as an upper-bound demand. Based on this per-application analysis, we design an efficient server resource management procedure to solve subproblems \textbf{SP1} and \textbf{SP2} for $M$ kinds of applications under the sufficient-resource assumption, as shown in Algorithm~\ref{singleCluster}. The constrained intra-node allocation under (\ref{ys.p.1})-(\ref{ys.p.2}) is handled by the container-based scheme discussed in the following subsection.

\subsection{Container-based Resource Management Scheme}
\label{sec:CRMS}
Algorithm~\ref{singleCluster} solves the relaxed subproblems \textbf{SP1} and \textbf{SP2} under sufficient-resource assumptions for each application, yielding an ideal configuration $c_i^*=(r_i^{cpu*},r_i^{mem*},N_i^*)$ that represents the upper-bound demand of application~$i$. However, these per-application configurations cannot be directly deployed on a resource-limited edge server because the global CPU and memory constraints (\ref{ys.p.1})-(\ref{ys.p.2}) may be violated. To address this limitation, we design a \emph{Container-based Resource Management Scheme} (CRMS) that redistributes resources across heterogeneous applications while maintaining global feasibility. CRMS exploits container elasticity to reallocate CPU and memory resources, reducing them for applications with low sensitivity to resource changes and assigning more to latency-sensitive ones, thus improving overall utility under constrained conditions.

\begin{algorithm}[h]
	\caption{Efficient Server Resource Management in Sufficient Resource Condition}
	\label{singleCluster}
	\renewcommand{\algorithmicrequire}{\textbf{Input:}}
	\renewcommand{\algorithmicensure}{\textbf{Output:}}
	\begin{algorithmic}[1]
		\REQUIRE $\overline{R^{cpu}}$, $\overline{R^{mem}}$, $\alpha$, $\beta$, $\lambda_{i}$,$\kappa_{i,1},\kappa_{i,2},\kappa_{i,3}$, $r_{i}^{min}$, $r_{i}^{max}$ 
		\ENSURE $r_i^{{cpu}^*},r_{i}^{{mem}^*},N_{i}^{*}$
		\FOR{$i= 1 \to M$}
		\STATE Obtain $r_i^{{cpu}^*},r_i^{{mem}^*}$ by Scipy.Optimize;
		\STATE Derive $\mu_{i}^{*}$ by substituting $r_i^{{cpu}^*},r_i^{{mem}^*}$ into Eq.(\ref{eq.uxd});
		\STATE Let $l=\left\lceil \frac{\lambda_i}{\mu_i^*} \right\rceil$, $r=\min \{\overline{R^{cpu}}/{r_i^{cpu}}^*,\overline{R^{mem}}/{r_i^{mem}}^*\}$;
		\WHILE{$l < r$}
		\STATE $lmid=l+(r-l)/3$;
		\STATE $rmid=r-(r-l)/3$;
		\STATE Calculate $\Phi (lmid)$ and $\Phi (rmid)$ by Eq.(\ref{sp2target});
		\IF{$\Phi (lmid)\le \Phi (rmid)$ }
		\STATE $r=rmid-1$;
		\ELSE
		\STATE $l=lmid+1$;
		\ENDIF
		\ENDWHILE
		\STATE $N_{i}^{*}=l$;
		\ENDFOR
	\end{algorithmic}
\end{algorithm}

\begin{figure}[h]
	\centering
	\begin{minipage}[b]{0.24\textwidth}
		\centering
		\includegraphics[width=\textwidth]{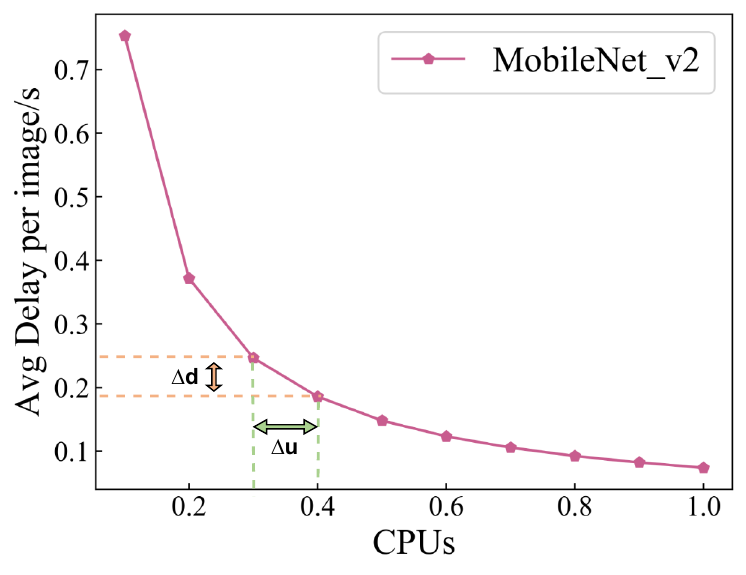}
		\\ % 换行符将 (a) 放到图片下方
		(a)
		\vspace{-1em}
	\end{minipage}
	\hfill
	\begin{minipage}[b]{0.24\textwidth}
		\centering
		\includegraphics[width=\textwidth]{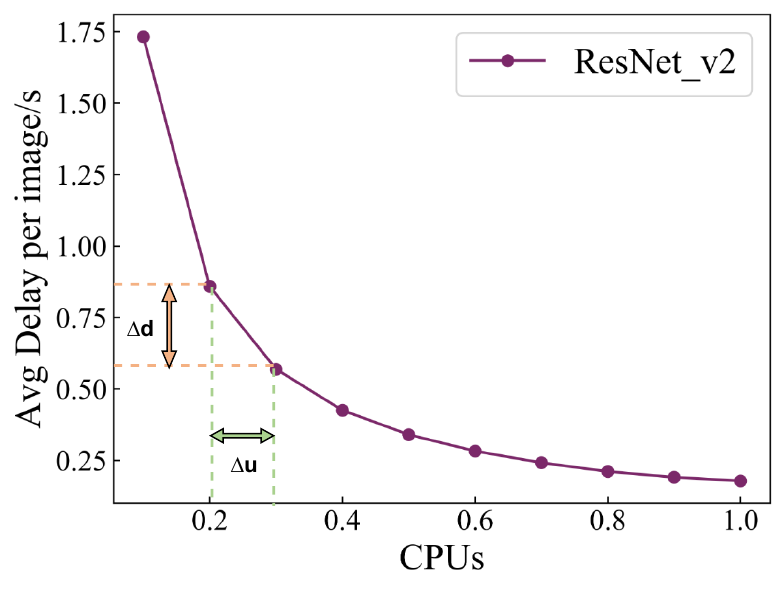}
		\\ % 换行符将 (b) 放到图片下方
		(b)
		\vspace{-1em}
	\end{minipage}
	\caption{(a) Container with small delay variation per unit CPU change. (b) Container with large delay variation per unit CPU change.}
	\label{fig:cpuScaleComparison}
\end{figure}

%To illustrate the efficiency of our resource reallocation scheme using CPU resource as an example (though it can be extended to other types of computational resources), we present the relationship between the CPU of application containers and their response latency, as shown in Fig.\ref{fig:cpuScaleComparison}. When containers are allocated a relatively large amount of CPU resources (as depicted in Fig.\ref{fig:cpuScaleComparison}(a)), reducing the CPU by $u$ units results in only a slight increase in processing delay. Conversely, when containers have fewer CPU resources allocated (as depicted in Fig.\ref{fig:cpuScaleComparison}(b)), the same reduction of $u$ units in CPUs leads to a significant increase in delay. Based on this observation, we aim to optimize server resource management by reallocating resources. Specifically, we intend to reduce the CPU  for applications similar to those in Fig.\ref{fig:cpuScaleComparison}(a), where a reduction in CPUs does not significantly impact delay. This allows us to allocate more CPU resources to applications that experience a substantial decrease in delay when their CPUs are increased, as shown in Fig.\ref{fig:cpuScaleComparison}(b).
To illustrate the motivation behind CRMS, Fig.~\ref{fig:cpuScaleComparison} depicts how container performance varies with allocated resources. When a container is allocated abundant CPU resources (Fig.~\ref{fig:cpuScaleComparison}(a)), reducing the CPU quota by $u$-units leads to only a marginal increase in latency. Conversely, when resources are scarce (Fig.~\ref{fig:cpuScaleComparison}(b)), the same $u$-unit reduction results in a significant latency increase. A similar trend is also observed for memory allocation, though the specific sensitivity pattern differs across applications. This observation underpins the CRMS design: resources are slightly reduced for delay-insensitive applications and reallocated to latency-sensitive ones, thereby improving overall system performance under resource constraints.

By effectively balancing per-application resource allocation, the scheme enhances the overall quality of service and ensures that critical applications receive sufficient computational resources. At this point, Problem~\textbf{P} can be transformed into Problem~\textbf{P1}, formulated as
\begin{flalign}
	\textbf{\text {P1:}}\min_{r_{i}^{cpu},r_{i}^{mem}}~ 
	&\sum\nolimits_{i}\!\left(\alpha W_{s}(N_i^*,\lambda_{i},\mu_{i})+\beta \frac{\Delta P(N_i^*)}{\lambda_{i}}\right),\\
	\text{s.t. }~ 
	&\text{Constraints.}(\ref{ys.p.1})-(\ref{ys.p.3})~\text{and}~(\ref{c:PNi}). \nonumber
\end{flalign}
Here $N_i^*$ represents the optimal container number derived by Algorithm~\ref{singleCluster} under sufficient resources assumption, while $r_{i}^{cpu}$ and $r_{i}^{mem}$ are reallocated CPU and memory resources. $\Delta P(N_i^*)$ denotes the incremental power consumption of $N_i^*$ containers for application $i$. 
To verify the existence of the optimal resource configuration for Problem~\textbf{P1}, we analyze the convexity of $U_i$ with respect to $r_{i}^{cpu}$ and $r_{i}^{mem}$ as follows.

\begin{theorem}\label{theorem.FixedN}
	$U_i$ is a convex function with respect to the resources $r_{i}^{cpu}$ and $r_{i}^{mem}$ when $N_{i}^*$ is given.
\end{theorem}

\begin{IEEEproof}
	Based on Constraint (\ref{c:PNi}), $\Delta P_i(N_i^*)$ is linear with respect to resource variables and therefore does not affect convexity. Thus, the convexity of $U_i$ depends on that of $W_{s}(N_i^*,\lambda_{i},\mu_{i})$ with respect to the resources $r_{i}^{cpu}$ and $r_{i}^{mem}$. According to \cite{zhang2020efficient}, in an M/M/N queuing system, it has been proven that $\frac{\partial W_s(N_i^*,\lambda_{i},\mu_{i})}{\partial \rho_i}>0$ and $\frac{\partial^2 W_s(N_i^*,\lambda_{i},\mu_{i})}{\partial (\rho_i)^2}>0$, where $\rho_i =\frac{\lambda_i}{N_i^*\mu_i}$. Using Eq. (\ref{eq.uxd}), $\rho_i$ is further represented as $\frac{\lambda_i \bar{x}_i d_i}{N_i^*}$, with $\lambda_i$, $\bar{x}_i$, and $N_i$ being constants in Problem \textbf{P1}. From this expression, we can derive that $\frac{\partial \rho_i}{\partial r_i^{cpu}}<0,\frac{\partial \rho_i}{\partial r_i^{mem}}<0$ and $\frac{\partial^2 \rho_i}{\partial (r_i^{cpu})^2}>0$ according to Theorem \ref{theorem.(2)}. Applying the chain rule, we can obtain the second derivative of $W_s(N_i^*,\lambda_{i},\mu_{i})$ with respect to $r_{i}^{cpu}$ and $r_{i}^{mem}$, i.e.,
	\begin{equation}
		\frac{\partial^2 W_s}{\partial\left(r_i^{cpu}\right)^2}=\frac{\partial^2 W_s}{\partial (\rho_i)^2} \left(\frac{\partial \rho_i}{\partial r_i^{cpu}}\right)^2 \\
		+\frac{\partial^2 \rho_i}{\partial\left(r_i^{cpu}\right)^2}  \frac{\partial W_s}{\partial \rho_i}>0, 
	\end{equation}
	\begin{equation}
		\frac{\partial^2 W_s}{\partial\left(r_i^{mem}\right)^2}=\frac{\partial^2 W_s}{\partial (\rho_i)^2} \left(\frac{\partial \rho_i}{\partial r_i^{mem}}\right)^2+\frac{\partial^2 \rho_i}{\partial\left(r_i^{mem}\right)^2}  \frac{\partial W_s}{\partial \rho_i}>0  
	\end{equation}
	and 
	\begin{equation}
		\frac{\partial^2 W_s}{\partial r_i^{cpu}\partial r_i^{mem}}=\frac{\partial^2 W_s}{\partial (\rho_i)^2} \frac{\partial \rho_i}{\partial r_i^{cpu}}\frac{\partial \rho_i}{\partial r_i^{mem}}+\frac{\partial^2 \rho_i}{\partial r_i^{cpu}\partial r_i^{mem}}  \frac{\partial W_s}{\partial \rho_i}.  
	\end{equation}
	
	From Hessian Matrix, we can derive that $\frac{\partial^2 W_s}{\partial\left(r_i^{cpu}\right)^2}\cdot \frac{\partial^2 W_s}{\partial\left(r_i^{mem}\right)^2}-(\frac{\partial^2 W_s}{\partial r_i^{cpu}\partial r_i^{mem}})^2>0$ holds, which means that ${W_{s}}(N_i^*,\lambda_{i},\mu_{i})$ is a convex function with respect to $r_{i}^{cpu}$ and $r_{i}^{mem}$. Therefore, $U_i$ is a convex function with respect to $r_{i}^{cpu}$ and $r_{i}^{mem}$ given $N_{i}^*$.
\end{IEEEproof}

According to Theorem~\ref{theorem.FixedN}, when the container count $N_i^*$ is given, Problem~\textbf{P1} is a convex optimization problem over CPU and memory allocations. This provides the theoretical basis for global resource reallocation in constrained edge servers. The optimal solution to \textbf{P1} can be efficiently obtained via convex solvers and, together with Algorithm~\ref{singleCluster}, forms the foundation of the proposed CRMS.

\begin{algorithm}
	\caption{Container-based Resource Management Scheme (CRMS)}
	\label{alg:CRMS}
	\begin{algorithmic}[1]
		\REQUIRE $\overline{R^{cpu}}$, $\overline{R^{mem}}$, $\alpha$, $\beta$, $\lambda_{i}$,$\kappa_{i,1},\kappa_{i,2},\kappa_{i,3}$, $r_{i}^{min}$, $r_{i}^{max}$
		\ENSURE $C$
		%		\FOR{$i= 1 \to M$}
		%		\STATE Calculate $c_i = ({r_i^{cpu}}^*,{r_i^{mem}}^*,N_i^*)$ by Algorithm \ref{singleCluster};
		%		\STATE $C \gets C \cup \{c_i\}$;
		%		\ENDFOR
		\STATE Compute $c_i^*=(r_i^{cpu*},r_i^{mem*},N_i^*)$ for each application via Algorithm~\ref{singleCluster};
		\STATE $C \gets C \cup \{c_i\}$;
		%		\STATE Calculate $U_p$ by Eq.(8) and Initialize $U_p' = U_p$;
		\IF{$\sum\limits_i N_i^* r_i^{cpu*} > \overline{R_{cpu}}$ or $\sum\limits_i N_i^* r_i^{mem*} > \overline{R_{mem}}$}
		\STATE Given $N_i' = N_i^*, \forall i$
		\STATE Solve Problem~\textbf{P1} to obtain $r_i^{cpu'},r_i^{mem'}$;
		\STATE Update $C = \{c_i', \forall i\}$ and Calculate $U_p$ by Eq.(\ref{eq:p});
		\ENDIF
		\WHILE{\TRUE}
		\STATE Set $\mathcal{O} \gets \emptyset$;
		\FOR{$i = 1 \to M$}
		\STATE Set $\hat{N_i} = N_i' - 1$;
		\STATE Solve Problem~\textbf{P1} for $\hat{C}=\{\hat{c_i},\forall i\}$;
		\STATE Calculate $\hat{U_p}$ by Eq.(\ref{eq:p});
		\STATE $\mathcal{O} \gets \mathcal{O} \cup \{(\hat{U_p}, \hat{C})\}$;
		\ENDFOR
		\STATE $U_p^*, C^* \gets \operatorname*{argmin}_{(\hat{U_p}, \hat{C}) \in \mathcal{O}} (\hat{U_p})$;
		\IF{$U_p^* < U_p$}
		\STATE Update $C \gets C^*$, $U_p \gets U_p^*$;
		\ELSE
		\STATE Break;
		\ENDIF
		\ENDWHILE
	\end{algorithmic}
\end{algorithm}

Initially, Algorithm~\ref{singleCluster} computes per-application configurations $c_i^*=(r_i^{cpu*},r_i^{mem*},N_i^*)$ under unconstrained conditions (lines~1--2), serving as upper-bound references for Problem~\textbf{P}. If these configurations exceed the server’s total CPU or memory capacities (\ref{ys.p.1})--(\ref{ys.p.2}), we fix $N_i^*$ based on Theorem~\ref{theorem.FixedN} and solve Problem~\textbf{P1} to reallocate CPU and memory resources (lines~3--7), obtaining feasible configurations $c_i'$. Then, the overall utility $U_p$ is computed via Eq.(\ref{eq:p}). Even if the initial configurations satisfy resource constraints, the upper-bound nature of Algorithm~\ref{singleCluster} may not guarantee global optimality. Hence, a greedy refinement is applied to iteratively decrease $N_i$ for each application, re-solving \textbf{P1} at every step (lines~8--22). If a reduced configuration yields a smaller $U_p$, the update is accepted; otherwise, the process terminates when no further improvement is possible.

\textbf{Complexity Analysis}: The computational complexity of CRMS arises from solving convex optimization problems using the Sequential Least Squares Programming (SLSQP) method in the Scipy.optimize library~\cite{song2022latency}. Algorithm~\ref{singleCluster} iteratively determines optimal CPU and memory configurations and the instance count for $M$ applications by solving a convex problem with two decision variables (CPU and memory). The per-application solve costs $O(k_1 \cdot 2^3)$ (where $k_1$ is the SLSQP iteration count), giving $O(M \cdot k_1 \cdot 2^3)$ across all applications. In addition, the ternary search used to refine $N_i$ contributes $O(M \cdot \log_3(\Delta) \cdot T_{\Phi})$, where $\Delta=\min\!\left\{\frac{\overline{R^{\text{cpu}}}}{r_i^{\text{cpu}*}},\,\frac{\overline{R^{\text{mem}}}}{r_i^{\text{mem}*}}\right\}-\left\lceil\frac{\lambda_i}{\mu_i^*}\right\rceil$ denotes the search range induced by queue stability and the server’s CPU/memory budgets, and $T_{\Phi}$ is the cost of evaluating $\Phi(\cdot)$ (including computing $W_s$ via Eq.~(\ref{Eq.mmn})). Hence, Algorithm~\ref{singleCluster} has complexity $O\!\left(M \cdot (k_1 \cdot 2^3 + \log_3(\Delta) \cdot T_{\Phi})\right)$. Algorithm~\ref{alg:CRMS} then proceeds in two phases: Phase~1 invokes Algorithm~\ref{singleCluster}, and Phase~2 refines allocations by repeatedly solving Problem~\textbf{P1} with $2M$ decision variables (CPU and memory for all applications), each with a per-solve cost of $O(k_2 \cdot (2M)^3)$, where $k_2$ is the number of iterations in the SLSQP solver. Over $K$ refinement iterations, solving $M$ instances of \textbf{P1} per iteration incurs $O(K \cdot M \cdot k_2 \cdot (2M)^3)$, so the total complexity of Algorithm~\ref{alg:CRMS} is $O\!\left(M \cdot (k_1 \cdot 2^3 + \log_3(\Delta) \cdot T_{\Phi}) + K \cdot M \cdot k_2 \cdot (2M)^3\right)$, dominated by $O(M \cdot k_1 \cdot 2^3)$ and $O(K \cdot M \cdot k_2 \cdot (2M)^3)$. 

In practice, CRMS operates in a quasi-dynamic manner: the allocator re-executes Algorithm~\ref{alg:CRMS} only when the monitor detects significant changes in the arrival rates $\{\lambda_i\}$ or in the active application mix, while container quotas remain fixed between optimization cycles. Thus, each optimization cycle incurs the polynomial-time complexity derived above. The re-optimization overhead scales with the number of applications and adaptation events, rather than with the per-request workload, making CRMS suitable for practical online use under global constraints (\ref{ys.p.1})–(\ref{ys.p.2}).

\section{Simulation and Experiments}
\label{sec:simulation}
\begin{figure*}[t]
	\centering
	\begin{minipage}[t]{0.32\textwidth}
		\centering
		\includegraphics[width=\textwidth]{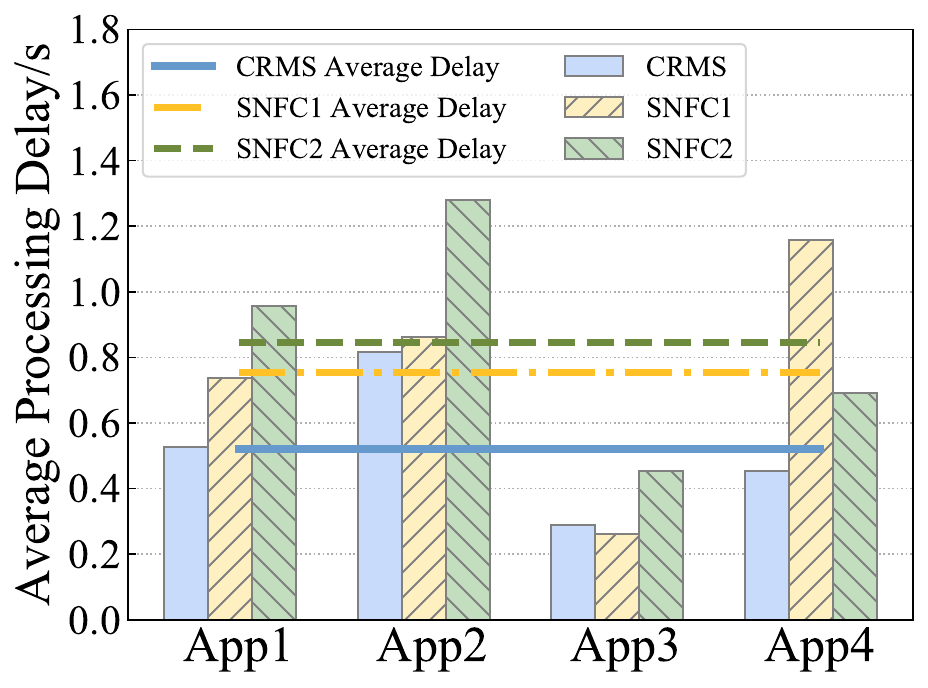}
		\vspace{-1em}
		\caption{Comparison of avg processing delay.}
		\label{diffConfDelay}
	\end{minipage}
	\hfill
	\begin{minipage}[t]{0.32\textwidth}
		\centering
		\includegraphics[width=\textwidth]{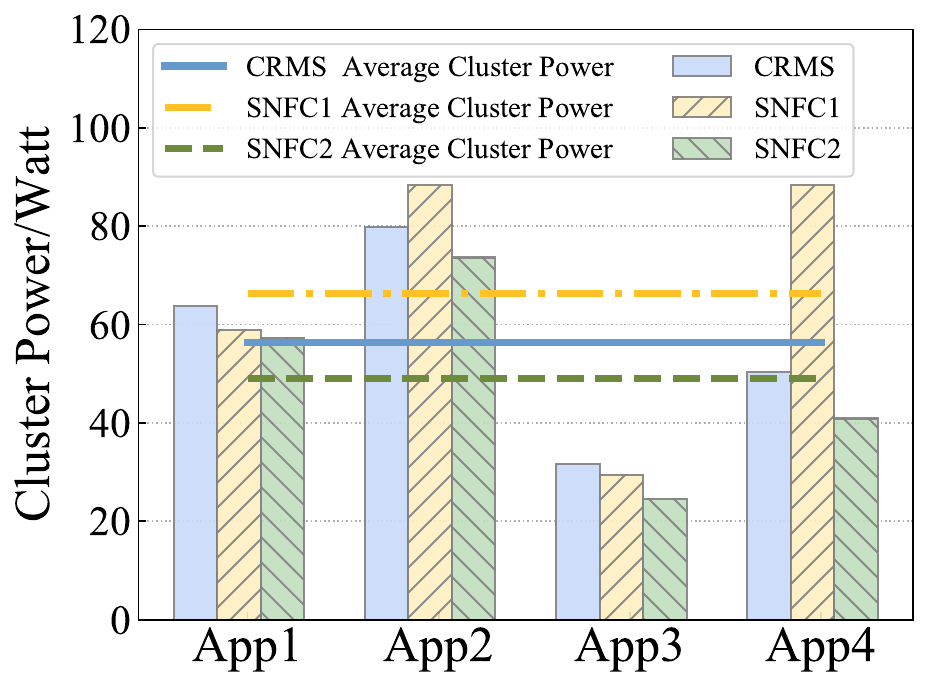}
		\vspace{-1em}
		\caption{Comparison of power consumption.}
		\label{diffConfPower}
	\end{minipage}
	\hfill
	\begin{minipage}[t]{0.32\textwidth}
		\centering
		\includegraphics[width=\textwidth]{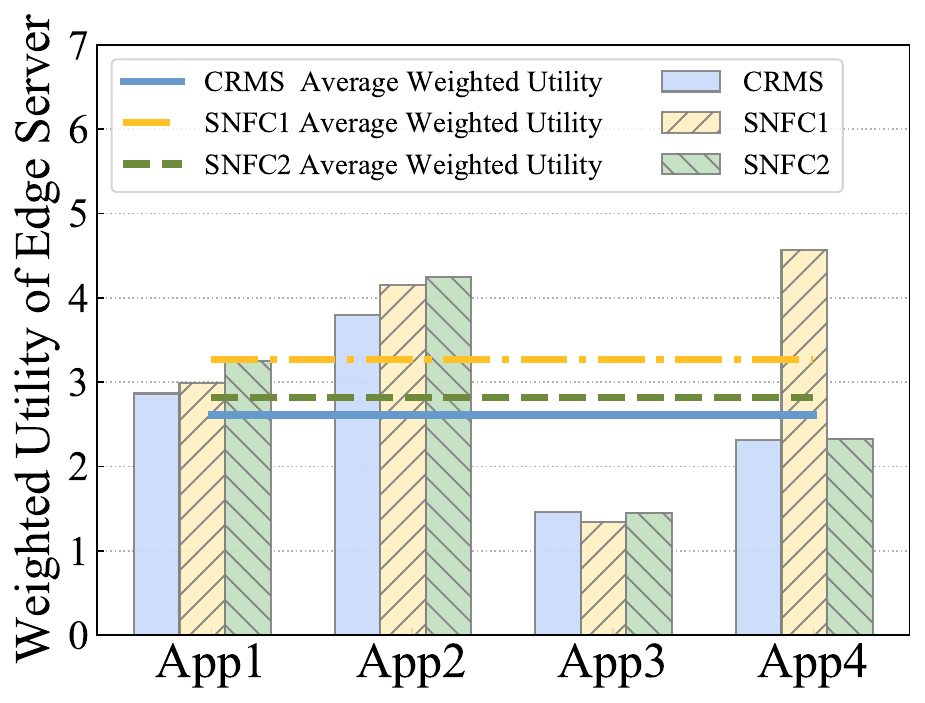}
		\vspace{-1em}
		\caption{Comparison of the edge server utility.}
		\label{diffConfUtility}
	\end{minipage}
	\vspace{-1em}
\end{figure*}
To evaluate the performance of the proposed scheme, we conduct extensive simulations from two perspectives. One is to evaluate the efficiency of the proposed scheme by comparing it to other existing schemes. The other is to analyze the impact of system parameters on performance in terms of processing delay and power consumption. 

The simulation environment is implemented in Python and executed on an Intel(R) Core(TM) i7-9700 processor. Simpy is used to simulate the task request process, while scipy.optimize is employed to solve the optimization problem. The experimental setup is defined as follows. Based on the profiling results in Section.\ref{sec:fitting}, we consider $M = 4$ applications, comprising three image classification tasks implemented using the ResNet\_v2, SE\_ResNeXt, and MobileNet\_v2 models, and one object detection task implemented using the SSD\_MobileNet\_v1 model. The minimum and maximum memory allocations for the four applications are set as $r^{min}=\{200,200,150,330\}$MB, $r^{max}=\{400,400,350,700\}$MB, respectively. The total available CPUs and memory resources range within $\overline{R^{cpu}}\in [28,38],\overline{R^{mem}}\in [6.5,11]$GB, respectively. For each application $i\in M$, the arrival rate $\lambda_{i}$ and the average number of images per request $x_i$ follow $\lambda_{i}\in[4,15],x_i\in [4,8]$. The task arrivals for each application follow a Poisson distribution, and the number of images per request is modeled using an exponential distribution. The weight parameters for processing delay and power consumption are set as $\alpha=1.4,\beta=0.2$, respectively. To simplify the description, we refer to ResNet\_v2, SE\_ResNeXt, MobileNet\_v2 and SSD\_ MobileNet\_v1 as APP1, APP2, APP3 and APP4, respectively.

\subsection{Performance Compared to Schemes}
As analyzed in Section \ref{sec:CRMS}, the amount of available resources on the edge server determines the outcome of the resource management strategy. Based on this observation, we conduct comparative experiments under both sufficient and constrained resource conditions. 

\textbf{Under sufficient resource conditions:} to highlight the impact of task heterogeneity on CPU and memory allocation, we compare our approach with a method called \textbf{SNFC}, which dynamically scales the number of pods while keeping their resource configurations fixed. Unlike our approach, existing methods such as \cite{9314906} often overlook the varying sensitivities of tasks to CPU and memory resources. This omission makes it difficult to configure resources optimally for diverse applications, resulting in degraded quality of service (QoS). For comparison, we set up two configurations of SNFC: \textbf{SNFC1} with $r^{cpu}=1.8$ and $r^{mem}=350$, and \textbf{SNFC2} with $r^{cpu}=1$ and $r^{mem}=r^{max}$. With $\lambda_i=6$ and $x_i=5,\forall i\in M$, the comparative results are presented below.

In Figs. \ref{diffConfDelay}-\ref{diffConfPower}, the CRMS algorithm achieves the lowest average processing delay across all four applications, while maintaining the second-lowest power consumption. This is due to its ability to dynamically allocate container and CPU resources according to the specific demands of each application. Although CRMS leads to slightly higher power consumption compared to SNFC2, it achieves a more effective trade-off between resource utilization and performance, particularly in reducing processing delays. According to Theorem \ref{theorem.(1)}, the processing delay exhibits a convex relationship with respect to CPU and memory resources. This means that while increasing both CPU and memory resources proportionally can help reduce latency, it does not necessarily result in optimal resource allocation. Therefore, simply increasing the number of containers may not always minimize the processing delay, as demonstrated in Fig. \ref{diffConfDelay}. Additionally, Fig. \ref{diffConfUtility} shows the edge utility, defined as the weighted average of processing delay and power consumption. CRMS achieves the lowest edge utility, indicating that it offers the best overall balance between delay reduction and power efficiency, outperforming both SNFC1 and SNFC2 in terms of overall system performance.

\begin{figure}[hbp]
	\centering
	\begin{minipage}[b]{0.24\textwidth}
		\centering
		\includegraphics[width=\textwidth]{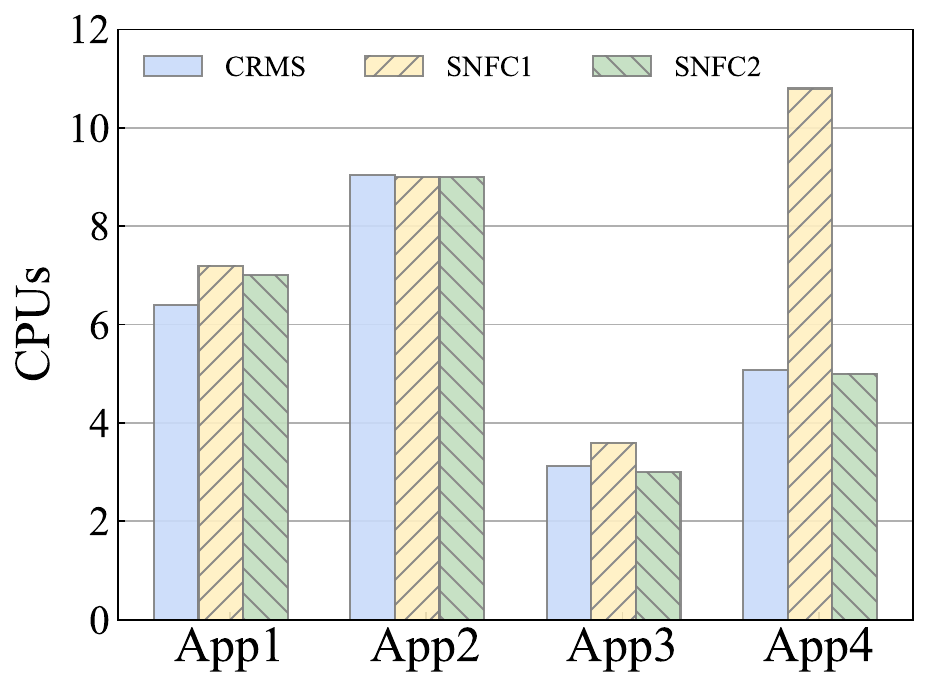}
		%		\vspace{-1em} % 调整图片和标题之间的间距
		\caption{Comparison of CPU usage.}
		\label{diffConfCPU}
	\end{minipage}
	\hfill
	\begin{minipage}[b]{0.24\textwidth}
		\centering
		\includegraphics[width=\textwidth]{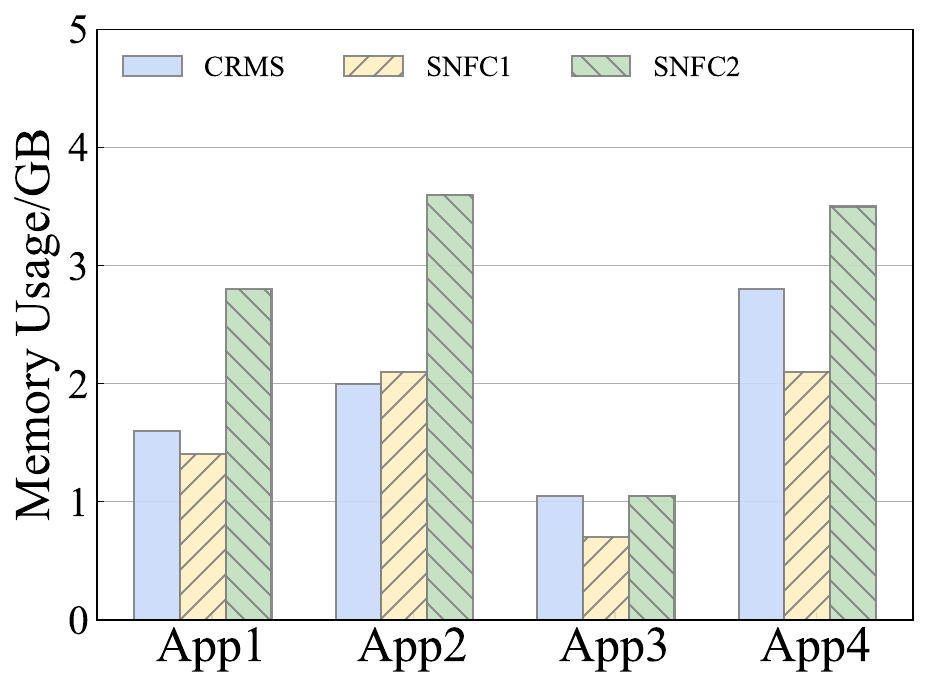}
		%		\vspace{-1em} % 调整图片和标题之间的间距
		\caption{Comparison of Mem usage.}
		\label{diffConfMEM}
	\end{minipage}
	\vspace{-1em}
\end{figure}

Given the container resource bounds defined by $r^{min}$ and $r^{max}$, SNFC1 exhibits a different behavior. While it satisfies the memory requirements for APP3, it struggles with APP2 and APP4 due to its limited memory allocation. APP2 is highly sensitive to memory availability, and even a slight undersupply of memory leads to significant increases in delay. In such cases, SNFC1 must allocate additional containers to compensate, thereby increasing both memory consumption and processing delay. In contrast, although APP4 is less sensitive to memory fluctuations, it requires a relatively large memory allocation to maintain efficient performance. When memory allocation is insufficient, APP4 experiences elevated delays. To meet performance demands, SNFC1 needs to open a larger number of containers, which results in increased CPU usage and higher power consumption, as shown in Fig. \ref{diffConfPower}. The higher CPU usage observed in SNFC1, particularly for APP4, reflects inefficient resource utilization, as illustrated in Fig. \ref{diffConfCPU}.

SNFC2 adopts a different strategy by allocating sufficient memory to all applications, thereby avoiding the performance degradation observed in memory-sensitive applications such as APP2. This sufficient memory provisioning prevents significant performance degradation. Consequently, SNFC2 exhibits the highest memory usage, as shown in Fig. \ref{diffConfMEM}. However, the CPU allocation per container in SNFC2 is slightly insufficient, leading to increased delays across all applications, as depicted in Fig. \ref{diffConfDelay}. Due to the linear relationship between CPU usage and power consumption, SNFC2 achieves lower power consumption compared to CRMS, as depicted in Fig. \ref{diffConfPower}. However, the under-provisioned CPU resources compel SNFC2 to deploy a larger number of containers for compensation, thereby reducing overall resource efficiency. Although SNFC2 achieves better power efficiency, its increased number of containers results in excessive memory usage and inefficient CPU utilization, highlighting an imbalance in resource allocation.

In contrast, CRMS dynamically adjusts both CPU and memory resources according to the specific needs of each application, ensuring a better trade-off between delay reduction and resource efficiency. Figs. \ref{diffConfCPU} and \ref{diffConfMEM} highlight how CRMS minimizes both CPU and memory overhead compared to the inefficiencies observed in SNFC1 and SNFC2. This dynamic resource allocation mechanism enables CRMS to achieve the lowest edge utility, demonstrating its effectiveness in managing heterogeneous application demands while optimizing both performance and resource utilization.

\begin{figure*}[t]
	\centering
	\begin{minipage}[t]{0.32\textwidth}
		\centering
		\includegraphics[width=\textwidth]{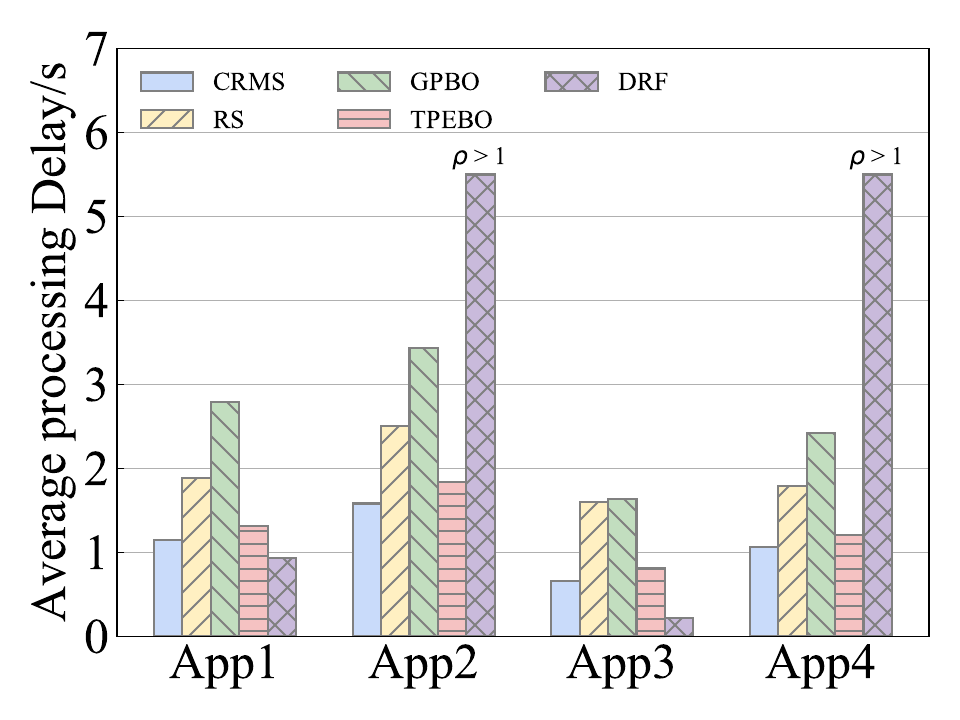}\vspace{-6pt}
		\caption{Comparison of processing delay.}
		\label{diffMethodDelay}
	\end{minipage}
	\hfill
	\begin{minipage}[t]{0.32\textwidth}
		\centering
		\includegraphics[width=\textwidth]{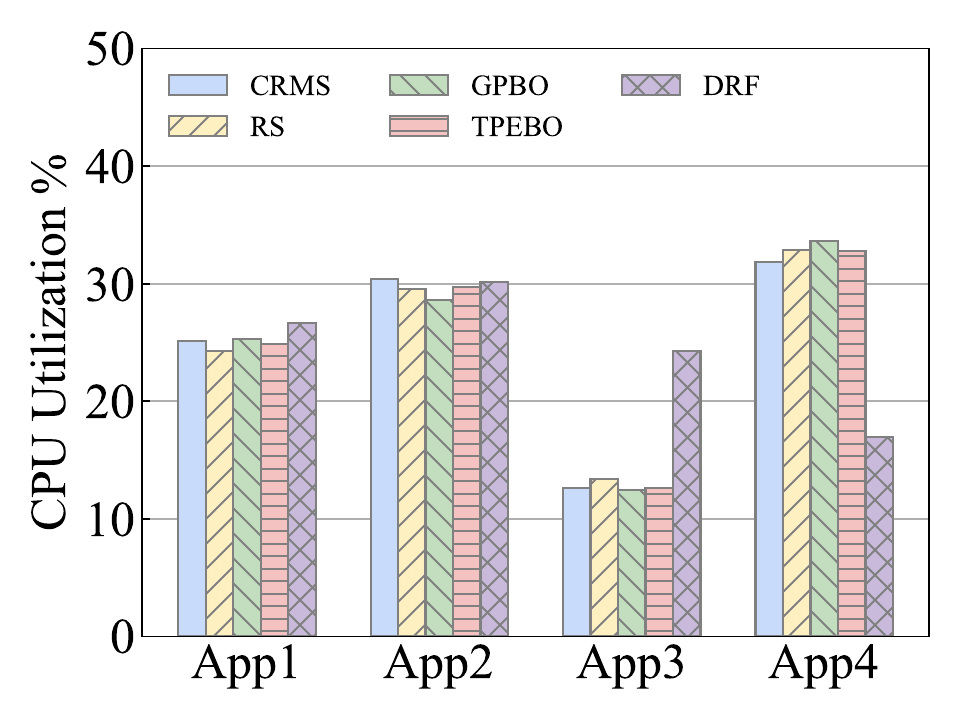}\vspace{-6pt}
		\caption{Comparison of CPU utilization.}
		\label{diffMethodCPU}
	\end{minipage}
	\hfill
	\begin{minipage}[t]{0.32\textwidth}
		\centering
		\includegraphics[width=\textwidth]{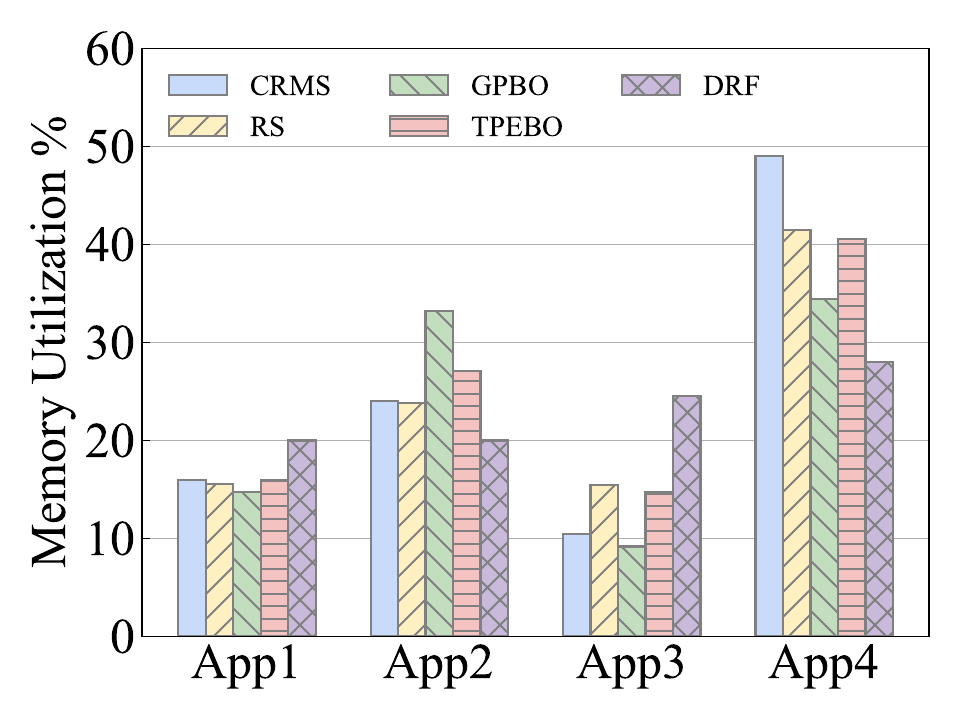}\vspace{-6pt}
		\caption{Comparison of memory utilization.}
		\label{diffMethodMEM}
	\end{minipage}
	\vspace{-1em}
\end{figure*}
\textbf{Under constrained resource conditions:} we evaluate the effectiveness of CRMS’s elastic scaling mechanism by comparing it with other baseline methods, given fixed total server resources. The comparison includes the following schemes:

\begin{itemize}
	\item Random Search (\textbf{RS}) \cite{bergstra2012random}: A baseline hyperparameter optimization method that samples configurations randomly from the parameter space, offering a straightforward benchmark for evaluating advanced algorithms.
	
	\item Gaussian Process Bayesian Optimization (\textbf{GPBO}) \cite{akhtar2020cose}: Builds a probabilistic model using Gaussian processes to identify promising configurations with fewer evaluations, suitable for costly optimization tasks such as container resource allocation.
	
	\item Tree-structured Parzen Estimator Bayesian Optimization (\textbf{TPEBO}) \cite{yu2023faasdeliver}: Separately models good and bad configuration distributions to guide the search process. It is well-suited for our tree-structured problem involving container count selection and CPU/memory allocation.
	
	\item Dominant Resource Fairness (\textbf{DRF}) \cite{meskar2022fair}: Ensures fair multi-resource allocation (e.g., CPU and memory) by equalizing dominant resource shares across tasks, promoting balanced resource usage in containerized environments.
\end{itemize}

The performance of the proposed CRMS scheme is evaluated against these algorithms to demonstrate its effectiveness in jointly optimizing container counts and resource allocations, particularly in heterogeneous and resource-constrained environments. From Eq.(\ref{eq:Cpu_usage}), the power consumption can be regarded as the constant given the total CPUs. Thus, we mainly compare the delay performance of CRMS, RPA, and DRF. By setting $\lambda_1=8,\lambda_2=7,\lambda_3=10,\lambda_4=15,x_i=5,\forall i\in M,\overline{R^{cpu}}=30,\overline{R^{mem}}=10$GB, the comparative results are shown as follows.

As shown in Fig. \ref{diffMethodDelay}, the CRMS, RS, GPBO, and TPEBO algorithms all successfully find feasible resource allocation schemes under constrained resource conditions. Notably, CRMS achieves the lowest average latency, reducing it by 43\%, 57\%, and 14\% compared to the other methods, respectively. Figs. \ref{diffMethodCPU}-\ref{diffMethodMEM} further demonstrate that CRMS handles the varying sensitivities of heterogeneous tasks to CPU and memory more effectively. This advantage arises because GPBO and TPEBO, constrained by the limited number of iterations, may not sufficiently explore the large parameter space, thereby making it challenging to converge on the optimal resource allocation. RS, by contrast, performs a random search across the entire parameter space without leveraging prior knowledge, which decreases the likelihood of finding an optimal solution.

The proposed CRMS starts with an initial resource allocation derived under unconstrained conditions and then iteratively adjusts the CPU and memory allocations for containers using a greedy approach to refine the configuration with fewer iterations. In comparison, the DRF algorithm prioritizes resources for applications with high demands, leading to resource shortages for other applications. This imbalance results in the task queues for APP2 and APP4 exceeding their capacity ($\rho>1$), indicating that these queues cannot be maintained within stable bounds. This comparison with DRF further underscores the importance of considering task heterogeneity in resource allocation strategies. Neglecting this factor may prevent certain applications from meeting their service requirements, often necessitating task offloading to other edge servers and thereby undermining the effective utilization of local computational resources. 

\begin{figure}[hbt]
	\centering
	\begin{minipage}[t]{0.45\textwidth}
		\centering
		\includegraphics[width=0.7\textwidth]{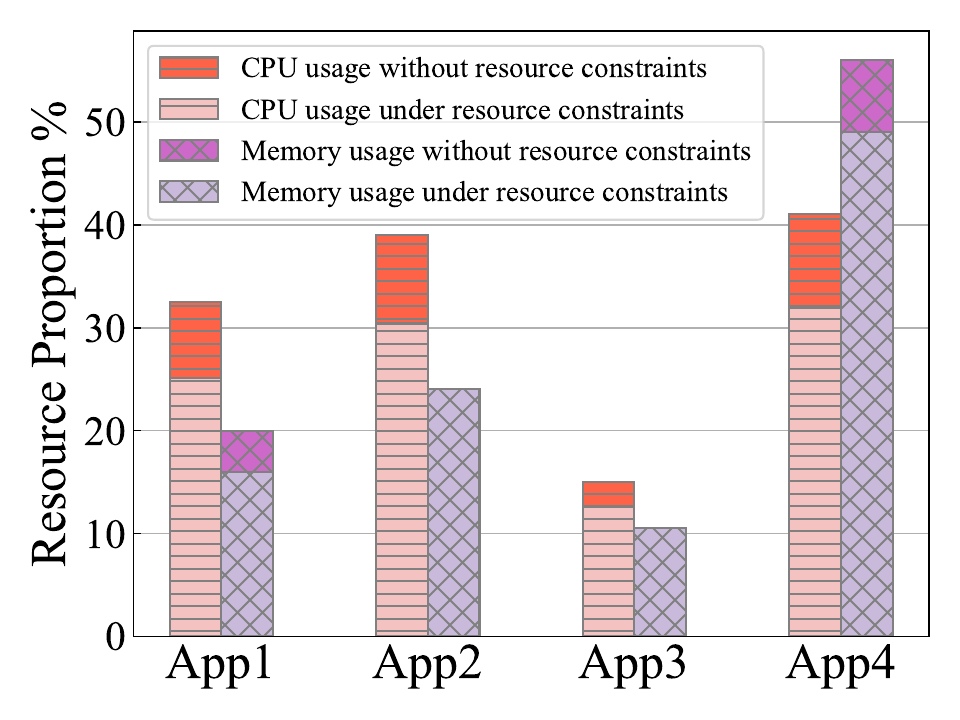}
	\end{minipage}\vspace{-6pt}
	\caption{Results of resource reallocation under resource constraints}
	\label{compressResource}
	\vspace{-1em}
\end{figure}

To verify the effectiveness of resource reallocation, we illustrate the changes in CPU and memory usage after incorporating resource constraints in Fig. \ref{compressResource}. The CRMS algorithm initially solves the resource allocation problem without constraints to obtain a preliminary solution, which is then iteratively refined under resource constraints. As a result, the system adjusts resource allocation based on the heterogeneous resource demands of different applications.

For CPU resources, all four applications undergo varying degrees of adjustment compared to the unconstrained scenario. The extent of CPU reallocation differs across applications, reflecting their sensitivity to computational resource constraints. In contrast, memory adjustments are primarily applied to APP1 and APP4, while APP2 and APP3 retain their original memory configurations. Additionally, the degree of memory reallocation varies between APP1 and APP4, highlighting the system’s ability to selectively compress memory resources where necessary while ensuring sufficient allocation for applications with higher memory demands.

\subsection{Analysis of System Parameters}
\begin{figure*}[t]
	\centering
	\begin{minipage}[b]{0.24\textwidth}
		\centering
		\includegraphics[width=\textwidth]{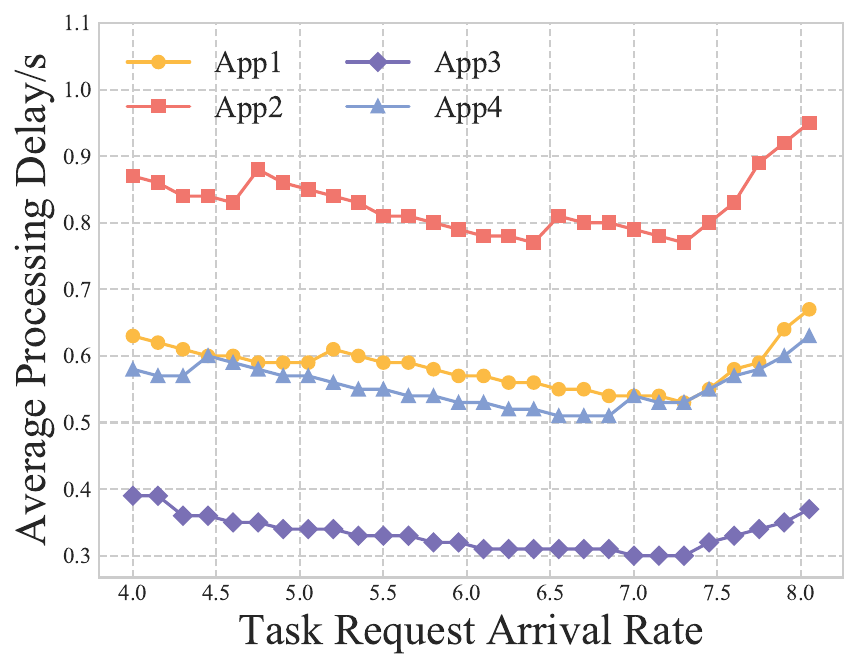}
		%		\vspace{-3pt}
		\caption{Processing delay with different task arrival rates.}
		\label{variableLamDelay}
	\end{minipage}
	\hfill
	\begin{minipage}[b]{0.24\textwidth}
		\centering
		\includegraphics[width=\textwidth]{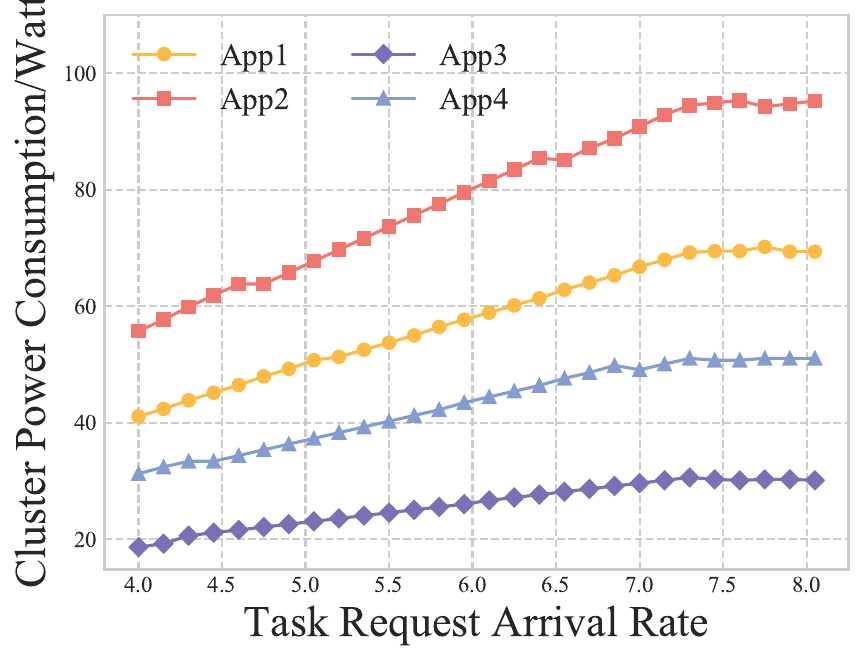}
		%		\vspace{-3pt}
		\caption{Power consumption with different task arrival rates.}
		\label{variableLamPower}
	\end{minipage}
	\hfill
	\begin{minipage}[b]{0.24\textwidth}
		\centering
		\includegraphics[width=\textwidth]{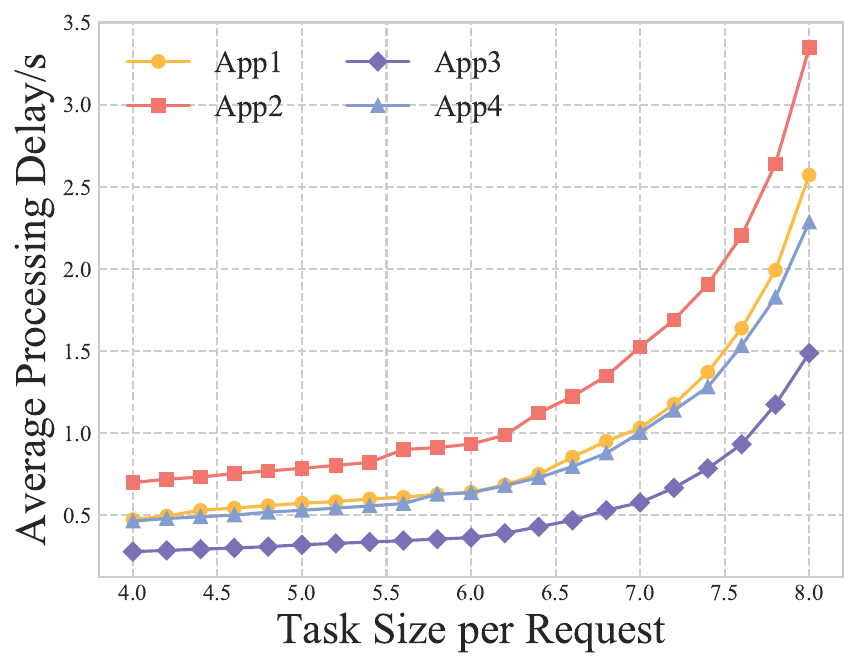}
		%		\vspace{-3pt}
		\caption{Processing delay under different request workloads.}
		\label{variableXDelay}
	\end{minipage}
	\hfill
	\begin{minipage}[b]{0.24\textwidth}
		\centering
		\includegraphics[width=\textwidth]{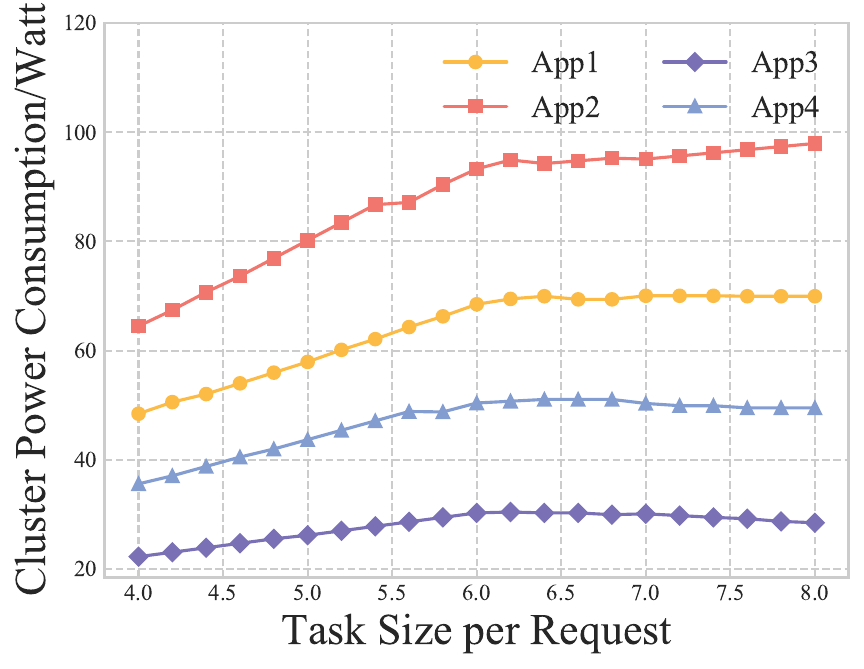}
		%		\vspace{-3pt}
		\caption{Power consumption under different request workloads.}
		\label{variableXPower}
	\end{minipage}

	\vspace{5pt} % 增加上下排之间的垂直间距
	
	\begin{minipage}[b]{0.24\textwidth}
		\centering
		\includegraphics[width=\textwidth]{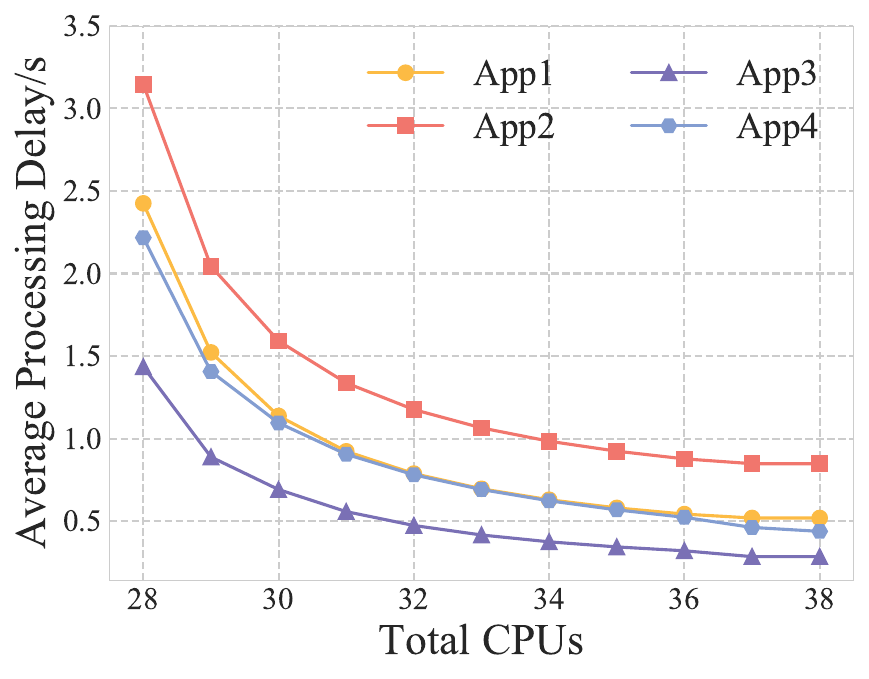}
		%		\vspace{-3pt}
		\caption{Processing  delay for different CPU resources.}
		\label{diffTotalCPU}
	\end{minipage}
	\hfill
	\begin{minipage}[b]{0.24\textwidth}
		\centering
		\includegraphics[width=\textwidth]{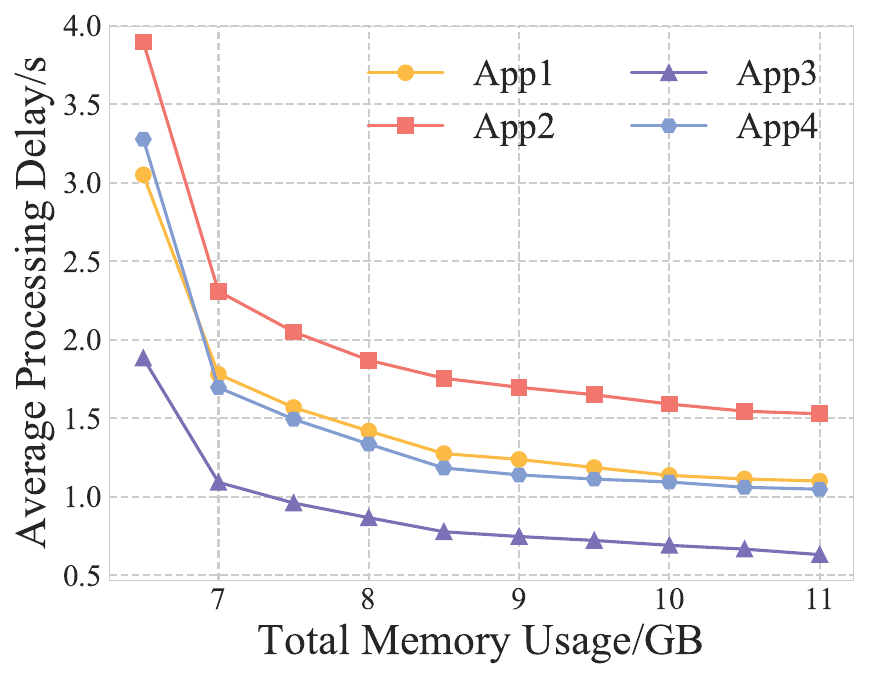}
		%		\vspace{-3pt}
		\caption{Processing delay for different memory resources.}
		\label{diffTotalMEM}
	\end{minipage}
	\hfill
	\begin{minipage}[b]{0.24\textwidth}
		\centering
		\includegraphics[width=\textwidth]{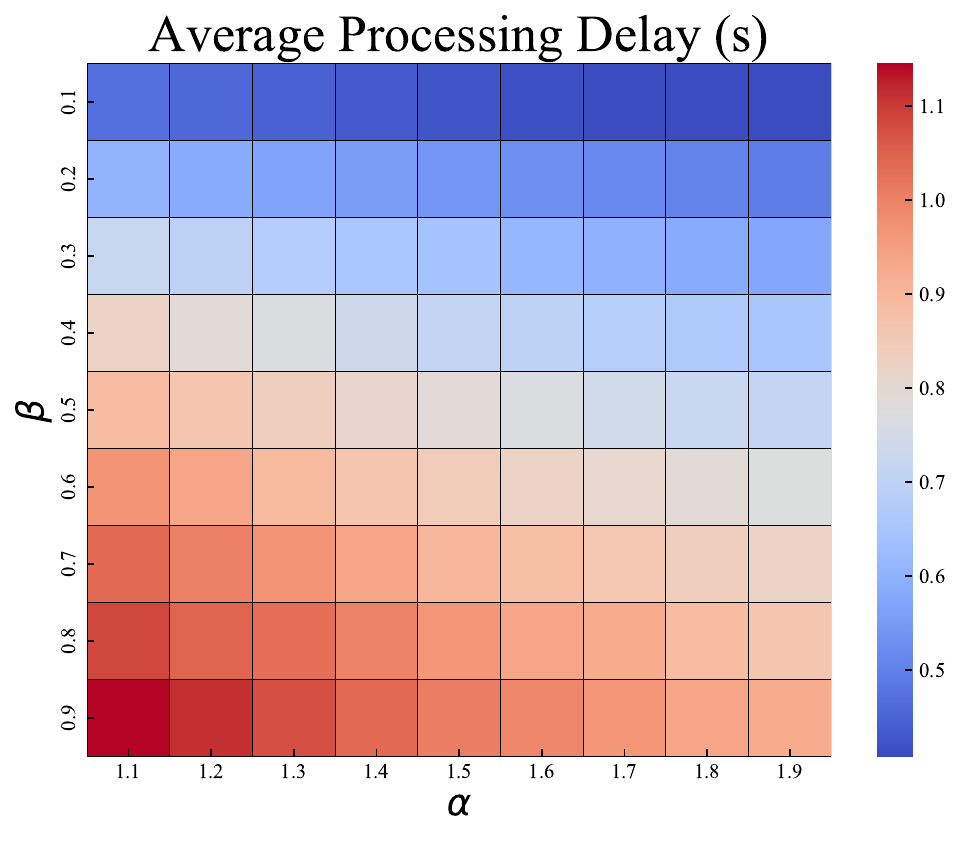}
		%		\vspace{-3pt}
		\caption{The impact of $\alpha$, $\beta$ on processing delay.}
		\label{heatmap_delay}
	\end{minipage}
	\hfill
	\begin{minipage}[b]{0.24\textwidth}
		\centering
		\includegraphics[width=\textwidth]{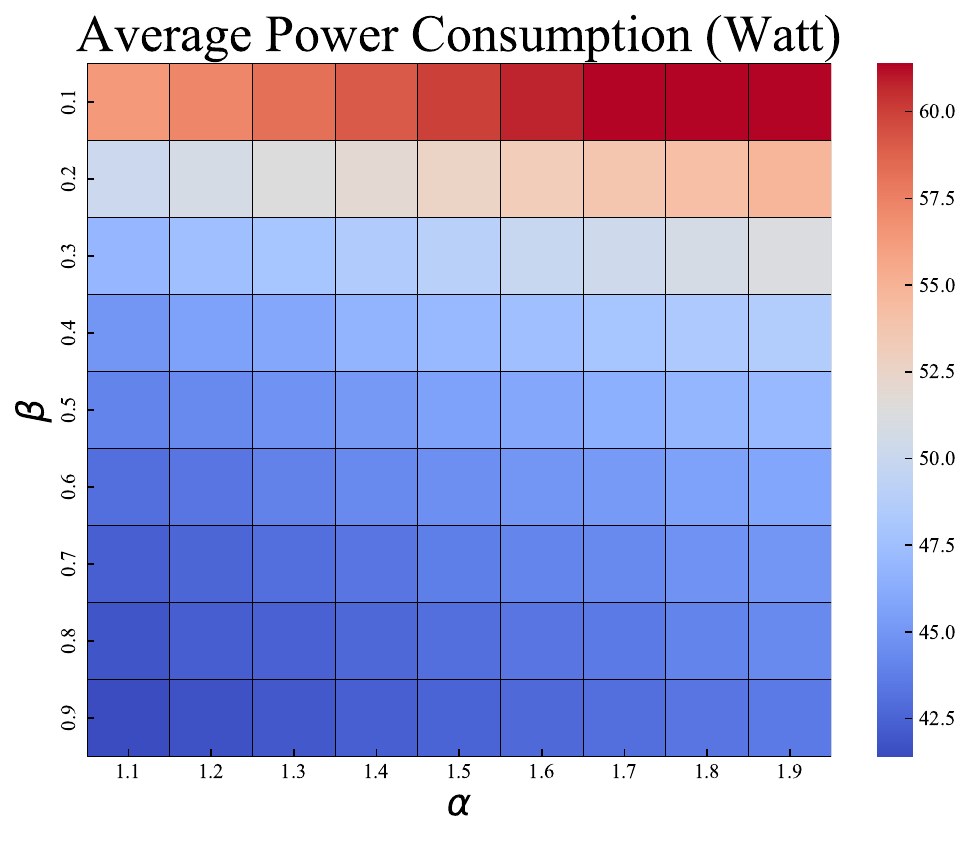}
		%		\vspace{-3pt}
		\caption{The impact of $\alpha$, $\beta$ on power consumption.}
		\label{heatmap_power}
	\end{minipage}
	\vspace{-1em}
\end{figure*}

Since the optimal results can be affected by system parameters, we analyze their impact from three aspects: 1) The task arrival rate $\lambda_{i}$, which reflects the performance of CRMS in processing delay and power consumption under varying workloads, as shown in Figs.\ref{variableLamDelay}-\ref{variableLamPower}; 2) The average number of images per request $x_i$, which characterizes the computational workload per request and influences the system’s service capacity and delay performance, as shown in Figs.\ref{variableXDelay}-\ref{variableXPower}; 3) The upper bounds of CPU and memory resources ($\overline{{{R}^{cpu}}},\overline{R^{mem}}$), which help assess the delay performance of CRMS under resource-constrained conditions through resource reallocation, as shown in Figs.\ref{diffTotalCPU}-\ref{diffTotalMEM}.

\textbf{The impact of $\lambda_{i}$:} As shown in Fig.~\ref{variableLamDelay}, the processing delay decreases as the task arrival rate ($\lambda_{i}$) increases, under the conditions of $\overline{R^{cpu}}=30$, $\overline{R^{mem}}=10\text{GB}$, and $x_i=5$. This trend occurs because the optimization objective defined in Equation (8) incorporates the time-averaged power consumption per request. As $\lambda_{i}$ increases, the relative weight of power consumption in the optimization objective diminishes, allowing the system to prioritize delay reduction. During this phase, CRMS dynamically adjusts resource allocation to optimize delay, resulting in an overall reduction in processing time. 

Additionally, periodic fluctuations in the delay curve are observed in Fig.\ref{variableLamDelay}, caused by CRMS activating additional containers to enhance parallelism when the CPU and memory resources of existing containers become insufficient. This reconfiguration temporarily reduces the resources allocated to each container, causing a slight increase in delay before the system stabilizes. As $\lambda_{i}$ approaches 7, server resources become constrained, intensifying competition among applications. At this stage, CRMS reallocates resources based on application sensitivity. Resource-intensive applications, such as APP2, experience increased delay due to higher contention, while lightweight applications like APP3 maintain relatively stable performance under constrained conditions. This mechanism allows CRMS to mitigate the impact of resource contention, delaying overall performance degradation. Fig.\ref{variableLamPower} shows that during the resource-sufficient phase, cluster power consumption increases steadily as more containers are activated. Beyond $\lambda_{i}=7$, the power consumption plateaus, indicating full utilization of server resources. At this point, CRMS redistributes resources among existing containers based on application sensitivity, ensuring efficient resource use while minimizing delay increases.

\textbf{The impact of $x_{i}$:} as shown in Fig.\ref{variableXDelay}, under the condition of $\overline{R^{cpu}}=30$, $\overline{R^{mem}}=10\text{GB}$, and $\lambda_{i}=6$, the processing delay of the four applications remains relatively stable as the request workload ($x_{i}$) increases, until $x_{i}$ exceeds approximately 6. During this phase, CRMS dynamically adjusts CPU and memory allocations for containers to handle the growing task size efficiently. However, when $x_{i}$ surpasses 6, the delay for all applications begins to rise significantly, indicating that server resources have reached their limit and resource contention intensifies. The trends in delay reveal the influence of resource sensitivity. For higher resource-demanding applications, such as APP2, the delay increases more steeply after saturation, whereas applications like APP3, with lower resource demands, experience a slower increase in delay. This disparity occurs because CRMS reallocates resources dynamically by compressing allocations for less resource-sensitive applications (e.g., App1, App3, and App4) to support applications like App2, which require more resources to maintain performance. Fig.\ref{variableXPower} shows a corresponding trend in power consumption. Initially, power consumption increases steadily with $x_{i}$ as additional containers are activated and resources are allocated to meet task demands. After $x_{i}$ reaches 6, power consumption reaches a plateau for most applications, reflecting the server's full utilization. Notably, App2’s power consumption continues to rise, demonstrating that CRMS prioritizes its resource adjustments to accommodate higher task sizes for resource-intensive applications, even under constrained conditions.

\textbf{The impact of $\overline{{{R}^{cpu}}}$ and $\overline{R^{mem}}$:} as shown in Fig. \ref{diffTotalCPU}, under the experimental conditions where the task arrival rates ($\lambda_{i}$) for the four applications are 8, 7, 10, and 15, and the task size per request ($x_{i}$) is fixed at 5, the processing delay decreases non-linearly as $\overline{{{R}^{cpu}}}$ increases, with $\overline{R^{mem}}$ fixed at 10GB. For resource-intensive applications such as App2, the reduction in processing delay is more pronounced due to their higher demand for computational resources. Increasing $\overline{{{R}^{cpu}}}$ allows the system to allocate more CPU resources to unit containers or activate additional containers, effectively reducing resource contention. In contrast, lighter applications like App3 exhibit a gentler delay reduction curve, reflecting their lower sensitivity to increases in CPU resources. Fig. \ref{diffTotalMEM} shows the variation in processing delay with $\overline{R^{mem}}$, while fixing $\overline{{{R}^{cpu}}}$ at 30. The processing delay also decreases non-linearly as memory resources increase. Applications with higher memory requirements, such as APP2 and APP4, show significant improvements in delay reduction as $\overline{R^{mem}}$ grows, due to the enhanced ability to buffer and process tasks in parallel. Conversely, App3 shows minimal delay reduction, as its lower memory demands make it less affected by increases in memory capacity.

\textbf{The impact of $\alpha$ and $\beta$:}  The heatmaps in Fig. \ref{heatmap_delay} and Fig. \ref{heatmap_power} demonstrate the combined effects of $\alpha$ and $\beta$ on average processing delay (in seconds) and average power consumption (in watts), highlighting their overall trends. From the first heatmap, the average processing delay increases with $\beta$ for all values of $\alpha$, while $\alpha$ itself plays a subtle role, with higher $\alpha$ values generally resulting in slightly reduced delays at the same $\beta$. This indicates that $\beta$ primarily governs the delay, with $\alpha$ introducing secondary adjustments. Conversely, the second heatmap shows that average power consumption decreases as $\beta$ increases, with higher $\alpha$ values further reducing power consumption, especially when $\beta$ is large. The trend suggests that higher $\alpha$ and $\beta$ values improve energy efficiency at the expense of increased processing delay, while lower values of $\alpha$ and $\beta$ result in reduced delay but higher power consumption. The balance between $\alpha$ and $\beta$ highlights their complementary roles, where $\beta$ predominantly dictates the trade-off between delay and energy consumption, and $\alpha$ modulates this effect. These trends emphasize the necessity of jointly tuning $\alpha$ and $\beta$ to meet application-specific requirements, offering a pathway to balance computational efficiency and energy savings.

\section{Conclusion}
\label{sec:conclusion}
%In this paper, considering the tasks with heterogeneous types have different delay performances in terms of CPU and memory resources, we establish a container-based fitting function to characterize the relationship among the processing delay, CPU, and memory resources. Based on this, we investigate a container-based resource management framework for edge computing with heterogeneous tasks,  where edge servers determine the optimal number of containers and the corresponding resource configuration for each type of task. To achieve optimal resource management for edge servers, we formulate the container-based resource configuration optimization problem to minimize the processing delay and power consumption simultaneously,  which is MINLP, and propose an efficient container-based resource management scheme based on theorem analysis. Experimental results show that CRMS has better performance in terms of delay and power consumption compared to others, under the consideration of the heterogeneity in task types and resource requirements.
This study addressed intra-node resource orchestration for heterogeneous edge workloads by integrating empirical latency profiling with a container-level performance model and a queueing-based delay formulation. The resulting joint latency–energy optimization was formulated as a mixed-integer nonlinear program and shown to be computationally intractable. To make the problem solvable in practice, we decomposed it into tractable convex subproblems and designed a two-stage container-based resource management scheme (CRMS) that jointly configures container counts and CPU/memory quotas under global resource constraints. The scheme exhibits polynomial-time complexity per optimization cycle and supports quasi-dynamic reallocation triggered by workload changes, thereby reconciling theoretical tractability with practical applicability on a single edge server.

Extensive simulations confirmed tangible improvements in both latency and energy efficiency over heuristic and search-based baselines across diverse heterogeneous workloads, validating the proposed model–algorithm co-design. While this work focused on intra-node optimization, the modeling and optimization framework is readily extensible to multi-node edge–cloud settings by incorporating inter-server bandwidth limits, routing decisions, and end-to-end service requirements, as well as temporal adaptation for time-varying demand.

% \vfill
\bibliographystyle{ieeetr}
\bibliography{refs}

\end{document}